\documentclass[superscriptaddress,twocolumn,secnumarabic,amssymb,nobibnotes,showpacs,aps,prm,citeautoscript,floatfix,usenames,dvipsnames]{revtex4-1}

\usepackage[nointegrals]{wasysym}
\usepackage{color}
\usepackage{tikz}
\usetikzlibrary{shapes}

\usepackage{titlesec}

\newcommand\emptystar{\raisebox{-.1em}{\tikz{\node[draw,scale=0.4, star,star point height=.7em]{};} }}

\usepackage{amsmath} 
\usepackage[english]{babel}
\usepackage{graphicx}
\usepackage{epstopdf}

\usepackage{array}
\usepackage{float}

\usepackage[T1]{fontenc}
\usepackage{mathptmx}

\usepackage[utf8]{inputenc}
\usepackage{lmodern}
\usepackage{textcomp}
\usepackage{xcolor}
\usepackage{microtype}

\usepackage{bm}

\usepackage{soul,xcolor}
\setstcolor{red}

\begin{document}

\title{van der Waals forces stabilize low-energy polymorphism in B$_2$O$_3$:\\ Implications for the {\em crystallization anomaly}}
\author{Guillaume Ferlat}
\affiliation{Sorbonne Université, MNHN, UMR CNRS 7590, IRD, IMPMC, F-75005 Paris, France}
\author{Maria Hellgren}
\affiliation{Sorbonne Université, MNHN, UMR CNRS 7590, IRD, IMPMC, F-75005 Paris, France}
\author{François-Xavier Coudert}
\affiliation{Chimie ParisTech, PSL University, CNRS, Institut de Recherche de Chimie Paris, 75005 Paris, France}
\author{Henri Hay}
\affiliation{Sorbonne Université, MNHN, UMR CNRS 7590, IRD, IMPMC, F-75005 Paris, France}
\author{Francesco Mauri}
\affiliation{Sorbonne Université, MNHN, UMR CNRS 7590, IRD, IMPMC, F-75005 Paris, France}
\affiliation{Dipartimento di Fisica, Universit\`a di Roma La Sapienza, Piazzale Aldo Moro 5, I-00185 Roma, Italy}
\author{Michele Casula}
\affiliation{Sorbonne Université, MNHN, UMR CNRS 7590, IRD, IMPMC, F-75005 Paris, France}

\date{\today}%

\begin{abstract}
The cohesive energies and structural properties of recently predicted - and never synthesized - B$_2$O$_3$ polymorphs are investigated from \emph{first principles} using density functional theory and high-accuracy many-body methods, namely, the random phase approximation and quantum Monte Carlo. We demonstrate that the van der Waals forces play a key role in making the experimentally known polymorph (B$_2$O$_3$-I) the lowest in energy, with many competing metastable structures lying only a few kcal/mol above. Remarkably, all metastable crystals are comparable in energy and density to the glass, while having anisotropic and mechanically soft structures. Furthermore, the best metastable polymorph according to our stability criteria has a structural motif found in both the glass and a recently synthesized borosulfate compound. Our findings provide new perspectives for understanding the B$_2$O$_3$ anomalous behavior, namely, its propensity to vitrify in a glassy structure drastically different from the known crystal.
\end{abstract}

\maketitle

\section{INTRODUCTION}

Diboron trioxide (B$_2$O$_3$) not only is the second most used component of industrial glasses after silica (SiO$_2$), but also a canonical network-forming system {\em per se} (see, e.g., Refs.~\cite{ferl_2015_1,wrig_2018_1} for reviews). The originality of the low-pressure B$_2$O$_3$ networks, either crystalline or vitreous, stems from their building blocks, which are bidimensional (2D) trigonal (BO$_3$) units. This is in contrast with most network formers, such as silica, based on three-dimensional (3D) tetrahedral units. Fully 3D networks are then formed by binding these rigid units through flexible cation-oxygen-cation bonds, which give B$_2$O$_3$ low-density structures, and a great potential for polymorphic transformations, under, e.g., high temperature or pressure. This is clearly reflected by various studies showing polyamorphic transformations in the glass~\cite{nich_2004_1,lee_2005_1,trac_2008_1,zeid_2014_1,lee_2018_1} and to a lesser extent in the liquid~\cite{braz_2010_1,alde_2015_1}.

However, our knowledge of the B$_2$O$_3$ crystalline forms remains very limited: up to now, only one low-pressure BO$_3$-based crystal (B$_2$O$_3$-I) has been experimentally characterized~\cite{gurr_70_1} in addition to a high-pressure phase (B$_2$O$_3$-II), based upon BO$_4$ tetrahedral units. The possible existence of another low-density polymorph, of unidentified structure, has been reported long ago~\cite{cole_35_1}, but the status of this report remains unclear since subsequent attempts failed to reproduce it. This is in sharp contrast with the rich polymorphism found in other oxide systems: in silica~\cite{picc_2000_1}, for instance, more than 20 low-pressure polymorphs (from coesite to zeolites) built upon the same basic units have been experimentally reported.

\begin{figure*}[t]
\includegraphics[width=0.62\linewidth]{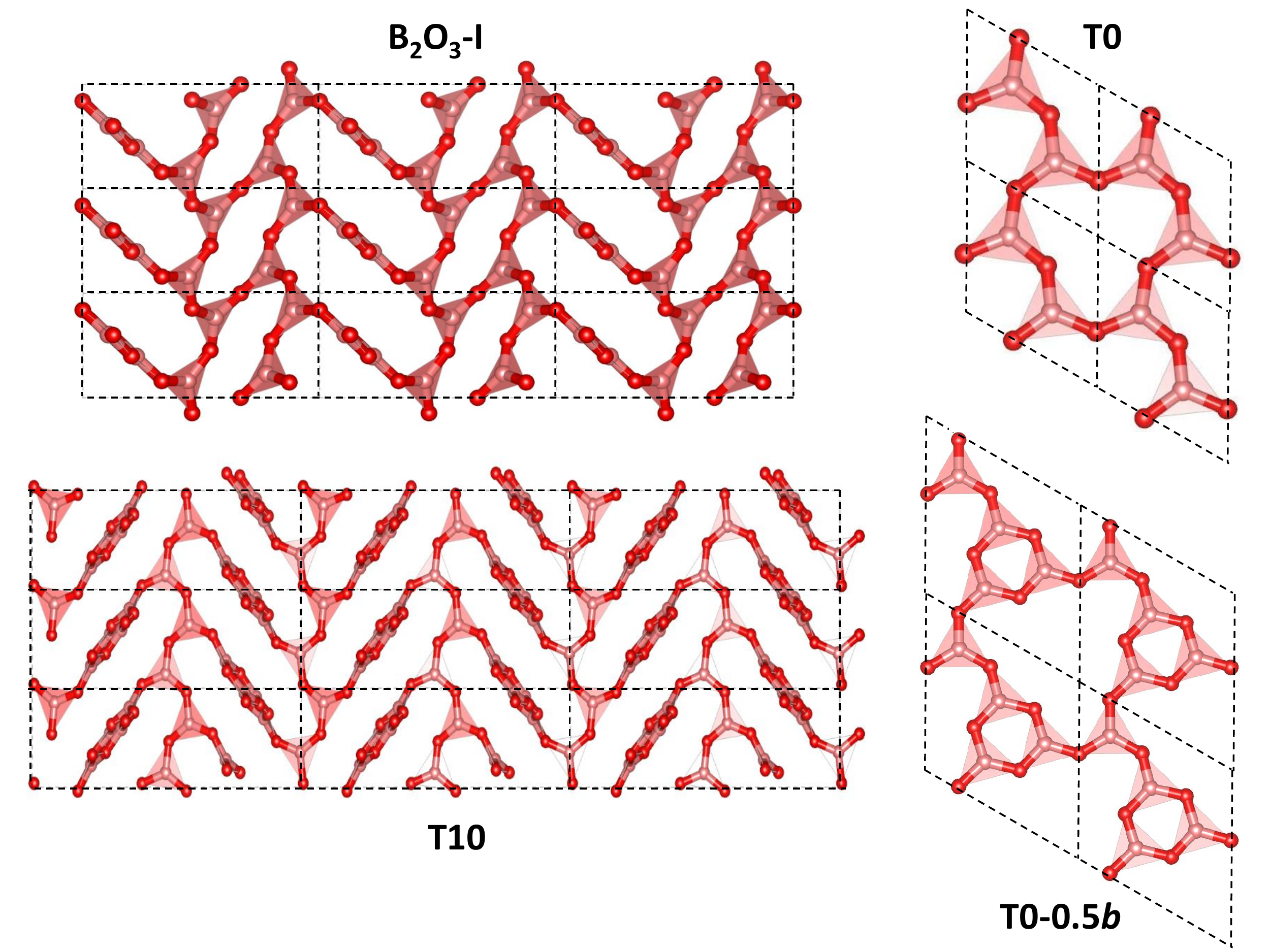}
\caption{Left: 3x3x3 supercell of B$_2$O$_3$-I and T10 seen in the ($a$,$c$) plane. Right: layers of T0 and T0-0.5$b$, the latter based on a mixed decoration of BO$_3$ triangle and boroxol units. All the polymorphs considered in this work are 3D networks with the exception of T0, T0-0.5$b$, and T0-$b$ which are layered structures.}
\label{fig_structures}
\end{figure*}

Another striking and very uncommon feature is the {\em abnormal structural dissimilarity} between the glassy (g-B$_2$O$_3$) and crystalline forms. In g-B$_2$O$_3$, about half of the BO$_3$ elemental bricks are arranged into superstructural units, i.e. threefold rings referred to as boroxol rings~\cite{youn_95_1,umar_2005_1}, fully absent from B$_2$O$_3$-I (see Fig.~\ref{fig_structures}). As a consequence, the glass density is considerably smaller ($\sim$ -30\%) than the B$_2$O$_3$-I one. Likely related~\cite{zano_2017_1} but yet not understood, is the extremely high glass-forming ability of B$_2$O$_3$, arguably the best glass former. Indeed, the B$_2$O$_3$-I crystallization has never been observed from ambient pressure liquid, even if seeded with germs for months. The synthesis requires cooling the liquid under pressure, or alternatively using chemical routes, a behavior which has been coined as the {\em crystallization anomaly}~\cite{uhlm_67_1}. 

Mostly inspired by the structural differences between g-B$_2$O$_3$ and B$_2$O$_3$-I, several theoretical works have predicted additional polymorphs~\cite{taka_2003_1,huan_2007_1,ferl_2012_1,clae_2013_1}.
In particular, a set of 25 new crystals~\cite{ferl_2012_1}, further complemented by two additional ones~\cite{clae_2013_1}, has recently been obtained using density functional theory (DFT). It spans a very narrow energy range (a few kcal/mol), with values comparable to, or even lower than, B$_2$O$_3$-I, therefore challenging it as the ground state. Hence, high-level theories are needed to go beyond the standard approximations (LDA, GGA) that have been used in all previous DFT works. As a matter of fact, a drastically different physical picture emerges from our high-level calculations as will be shown later.

In the current work, we employ accurate many-body methods such as quantum Monte Carlo (QMC) and the random-phase approximation (RPA) to provide a definitive answer to the relative stability between B$_2$O$_3$-I and a subset of predicted polymorphs. In addition, we present results for a large set of polymorphs at different levels of DFT which demonstrate a huge effect of the van der Waals (vdW) forces on both the structures and energetics, an effect that has been neglected in all previous studies~\cite{taka_2003_1,huan_2007_1,ferl_2012_1,clae_2013_1}. We definitely assess B$_2$O$_3$-I as the ground state while we introduce a novel polymorph (T0-0.5$b$), adapted from a recently synthesized borosulfate structure~\cite{daub_2015_1} which we reveal as the most stable among all putative structures. Its layers are decorated by triangle and boroxol units (Fig.~\ref{fig_structures}) in equal proportions (1:1), making it a close, albeit crystalline, structural approximant of the glass.

\section{METHODS}

Highly accurate QMC simulations have been carried out on three polymorphs (B$_2$O$_3$-I and two predictions T0 and T10). We used a Jastrow-Slater variational wave function, with Slater and Jastrow parts developed on a localized Gaussian basis set. The Jastrow factor contains correlation terms up to the four-body (electron-ion-electron-ion) form, able to capture van der Waals effects within variational Monte Carlo (VMC)~\cite{michele_benzene}. The wave function has been fully optimized (Slater orbitals together with Jastrow coefficients) by energy minimization~\cite{Umrigar2007}, starting from DFT-LDA generated one-body orbitals. Moreover, we have performed a complete structural relaxation for both cell parameters and internal coordinates at the VMC level~\cite{barborini2012structural}. Then, using relaxed VMC geometries, we have carried out lattice-regularized diffusion Monte Carlo simulations~\cite{michele_lrdmc,filippi_lrdmc} and a very accurate finite-size extrapolation~\cite{dagrada2016exact} in order to provide energies with accuracy better than 1.0 kcal/mol.
A careful convergence of all relevant criteria, i.e. basis set, geometry, finite-size effects and level of theory, is necessary, given the phase-space proximity of the B$_2$O$_3$ polymorphs. All QMC calculations have been performed using the TurboRVB code~\cite{sorella_rvb}. Further details can be found in the Supplemental Material (SM).

Calculations with the RPA were performed on a larger set of six polymorphs (B$_2$O$_3$-I, T0, T10, T3, T0-0.5$b$ and T0-$b$). The RPA is a state-of-the-art density functional approach based on many-body perturbation theory and the adiabatic connection fluctuation-dissipation formula~\cite{lang_75_1,gunn_76_1}. By including polarization diagrams to infinite order in the Coulomb interaction the RPA captures vdW forces~\cite{BN1,GRAPHRPA}. It also provides an accurate description of Hartree-Fock exchange, and an overall good description of correlation effects (at least as far as energy differences are concerned). RPA total energies were calculated with the VASP code~\cite{kres_96_1,kres_99_1}. Computational details can be found in the SM.

Dispersive interactions can also be added in a semiempirical form to standard DFT approximations such as in the DFT-D approaches: for instance, in D2~\cite{grim_2006_1} the vdW interactions coefficients ($C_6$) are fixed for a given atomic pair while in the Tkatchenko-Scheffler (TS) approach~\cite{tkat_2009_1}, they are calculated self-consistently according to the atomic neighborhood. At a more advanced level, vdW are accounted for in fully non-local functionals that include polarization effects from first-principles, such as in the vdW-DFT class of functionals~\cite{lee_2010_1,thon_2015_1}. In this work, we used representative functionals from these different levels, namely, D2~\cite{grim_2006_1}, TS~\cite{tkat_2009_1} [added on top of the Perdew-Burke-Ernzerhof (PBE) functional~\cite{perd_96_1}, using the CASTEP code~\cite{clar_2005_1}] and the recently derived DF-cx~\cite{thon_2015_1} vdW-DFT functional (using the Quantum Espresso package~\cite{QE-2009}). We checked explicitly that the use of different codes and pseudo-potentials does not affect the reported results (Fig. S1 of the SM).
We also report results from LDA and GGA (PBE) functionals as references. In particular, the comparison between GGA and dispersion-corrected schemes (D2 and TS) built upon the same GGA, allows one to probe straightforwardly the relevance of vdW interactions.
Thanks to the low computational cost of DFT, we studied a total of 27 polymorphs (those from Ref.~\cite{ferl_2012_1} supplemented with T0-0.5$b$), which contain up to 135 atoms per unit cell~\cite{ferl_2012_1}. 

Finally, we studied the mechanical properties via DFT-D2 calculations of second-order elastic constant tensors following a methodology described elsewhere~\cite{coud_2013_1,hay_thesis}. See the SM for all the details of the DFT calculations.

\section{RESULTS AND DISCUSSION}

\begin{figure}[t]
\includegraphics[height=4.5cm, width=8.5cm]{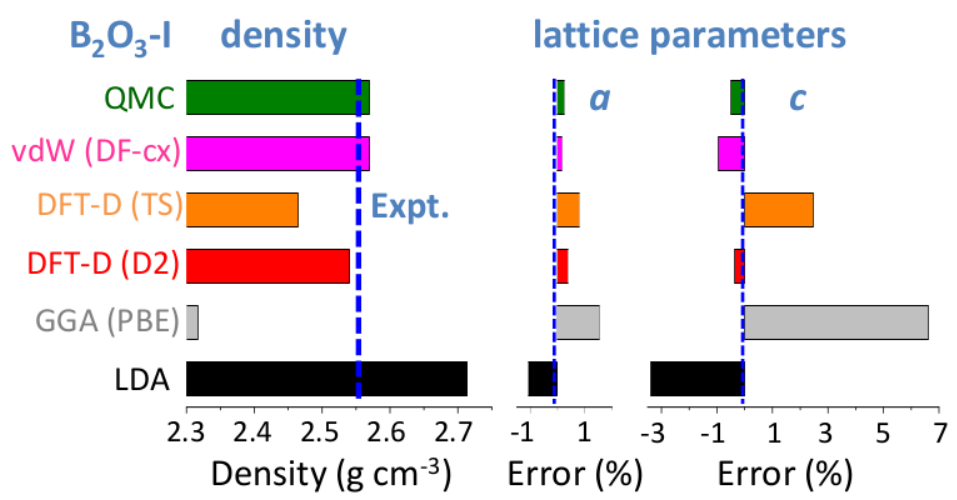}
\caption{Density and lattice parameters of B$_2$O$_3$-I from different {\em ab initio} schemes. Errors on lattice parameters are expressed relatively to the experimental values.}
\label{fig_benchmark}
\end{figure}


\begin{figure}[t]
\includegraphics[height=9.2cm, width=8.5cm]{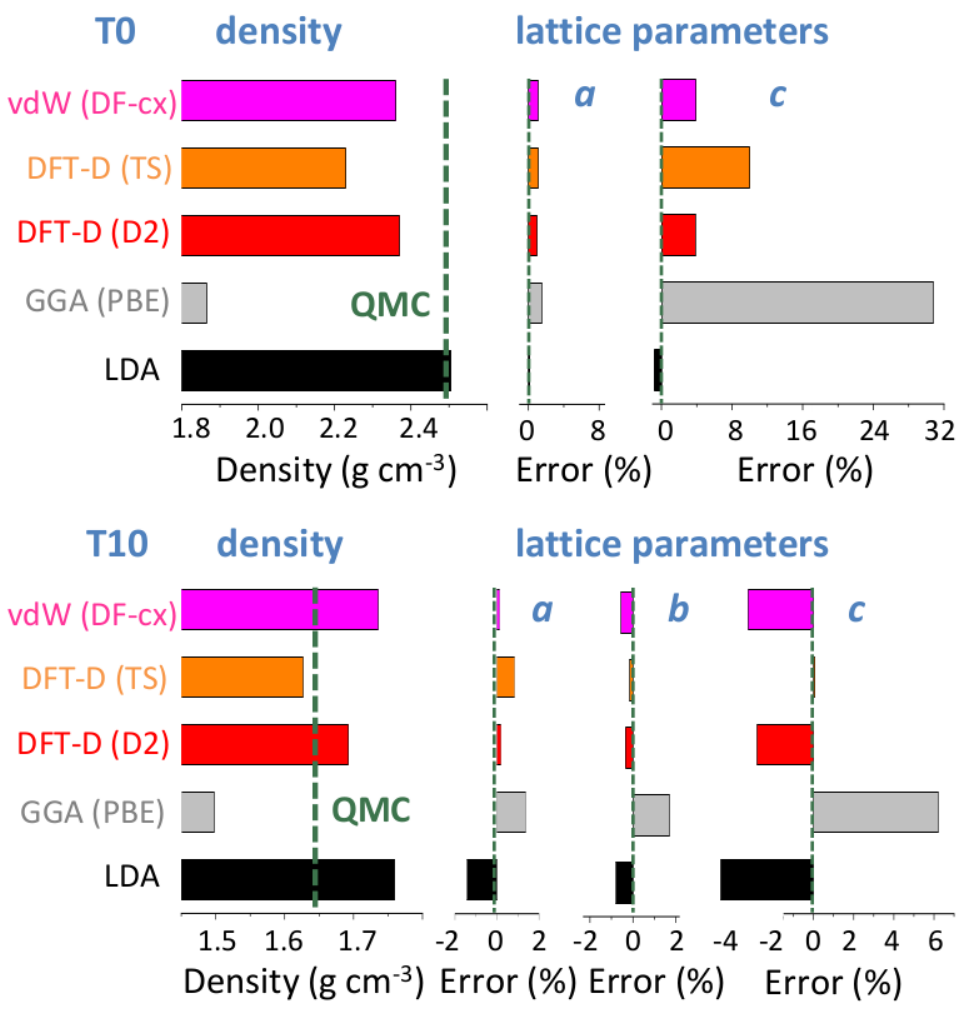}
\caption{Density and lattice parameters of T0 and T10 from different {\em ab initio} schemes. Errors on lattice parameters are expressed relatively to the QMC values.}
\label{fig_benchmark_T0-T10}
\end{figure}

We first assess the quality of the structural results obtained from the different schemes by relaxing the B$_2$O$_3$-I structure, and taking the experimentally known geometry as reference (Fig.~\ref{fig_benchmark}). Not surprisingly, the lattice parameters obtained with LDA are underestimated (and the density overestimated by +6\%) while the opposite is true for PBE (density error of -10\%). 
This reflects the well-documented tendencies~\cite{haas_2009_1,hay_2015_1} of LDA to overbind and of PBE to underbind. Note, however, that the size of these errors is large for a non-layered inorganic system, placing B$_2$O$_3$-I in the topmost range (95th percentile) of volume errors for inorganic materials of the Materials Project database~\cite{sun_2016_1}.
Interestingly, these deficiencies are largely cured not only by QMC (+0.5\% error on density) but also by all the vdW-corrected schemes used here, which show a systematic improvement.

Although B$_2$O$_3$-I is a fully connected 3D network made of strong interatomic bonds, the importance of the vdW corrections stems from the structural porosity, arising from {\em locally planar} regions - on the scale of a few building units - arranged in a {\em zig-zag} pattern thanks to the B-O-B angular flexibility (Fig.~\ref{fig_structures}).
This leads to a {\em softer} direction, perpendicular to the locally planar regions, and nearly parallel to the {\em c} direction, as reflected by larger errors in this lattice parameter (Fig.~\ref{fig_benchmark}). 
The existence of such a softer direction is found in all but two polymorphs (see also Fig.~\ref{fig_benchmark_T0-T10}).

\begin{figure}[t]
\includegraphics[height=6.cm, width=8.5cm]{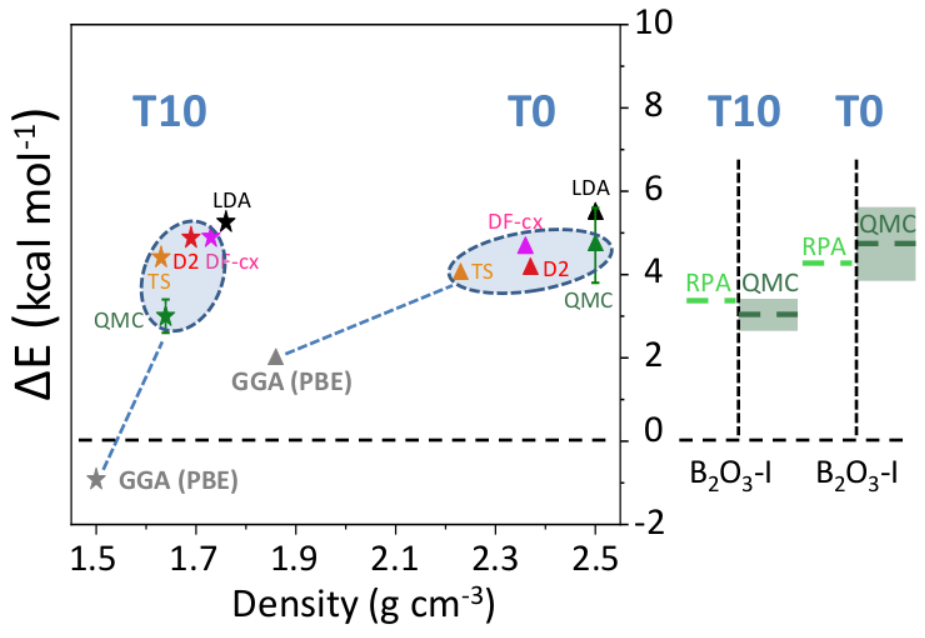}
\caption{Left: energies, relative to B$_2$O$_3$-I, from the different schemes for two predicted polymorphs, T0 (triangles) and T10 (stars). The blue-shaded ellipses encompass vdW-corrected results. Right: RPA and QMC energies compared. The green-shaded rectangles correspond to the QMC error bars.}
\label{fig_benchmark_2}
\end{figure}

We now focus on the densities and energies of two previously predicted~\cite{ferl_2012_1} polymorphs, namely T0 and T10 using QMC results as reference in the absence of experimental data. We report these results in Figs.~\ref{fig_benchmark_T0-T10} and \ref{fig_benchmark_2}. In the following, all crystals' energies are expressed with respect to the B$_2$O$_3$-I one.
Similarly to the experimentally known case, PBE severely underestimates the densities. This impacts the energies, which are also underestimated. 
Note that T10, which in the PBE original predictions~\cite{ferl_2012_1} had a slightly lower energy ($\sim -1$ kcal/mol) than B$_2$O$_3$-I, turns out to be metastable ($\sim +3$ kcal/mol).
On the contrary, the vdW-corrected DFT schemes perform reasonably well in both densities and energies. Moreover, the RPA energies agree well with QMC  (Fig.~\ref{fig_benchmark_2}, right panel). 

\begin{figure}[t]
\vskip -0.2cm
\includegraphics[width=0.59\linewidth]{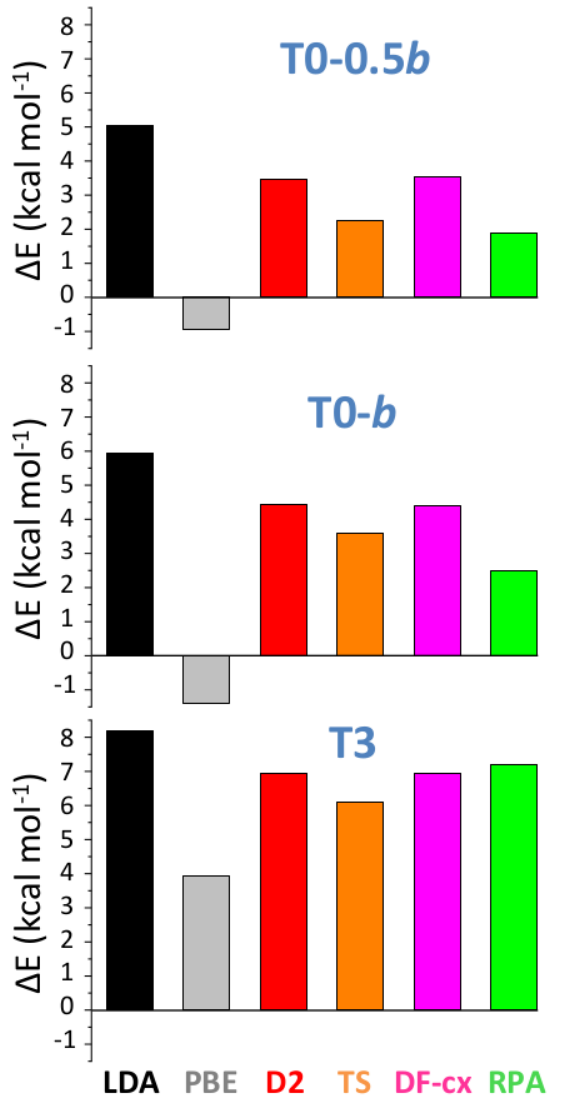}
\vskip -0.25cm
\caption{Energies, relative to B$_2$O$_3$-I (in kcal~mol$^{-1}$) obtained from the different {\em ab initio} schemes, including RPA, for three polymorphs (T0-0.5$b$, T0-$b$, and T3). The energies for T0 and T10 are reported in Fig.~\ref{fig_benchmark_2}.}
\label{fig_benchmark_3}
\end{figure}

We also calculated RPA energies for the T0-0.5$b$, T0-$b$, and T3 structures, shown in Fig.~\ref{fig_benchmark_3} together with those from the different DFT schemes. We note that all three vdW-corrected functionals yield essentially the same results, with a typical variability of 2 kcal/mol on the energies, and compare well with the more advanced RPA. This gives strong confidence in the overall picture.
A complete account for the full set of 27 B$_2$O$_3$ polymorphs, using the various DFT frameworks, is reported (Fig.~S2) in the SM.
In the following we shall take D2 as representative of the vdW-corrected functionals~\cite{BM_D2}, in order to make a thorough comparison with PBE, and show the impact of the dispersive interactions on the B$_2$O$_3$ polymorphism.

\begin{figure}[t]
\vskip -0.2cm
\includegraphics[height=5.4cm, width=8.5cm]{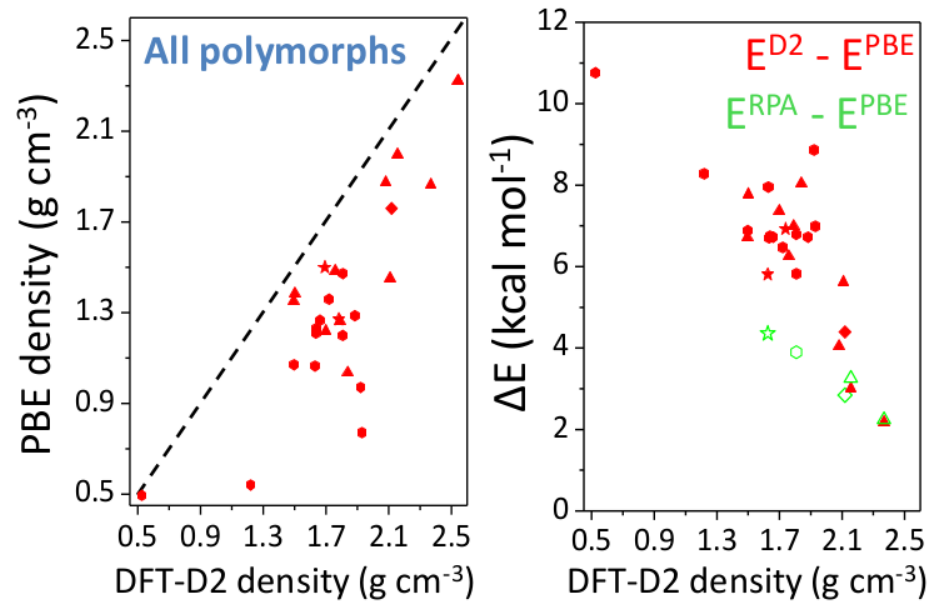}
\vskip -0.25cm
\caption{Corrections to the PBE density (left panel) and relative energy (right panel) brought back by the vdW-corrected schemes, using DFT-D2 (red) or RPA (green symbols).}
\label{fig_benchmark_4}
\end{figure}

Figure~\ref{fig_benchmark_4} highlights the density and energy corrections brought back by the vdW contributions. Being attractive, the vdW forces systematically provide denser structures (Fig.~\ref{fig_benchmark_4}, left panel), on average by 40\% and by up to 150\% for some polymorphs, such as T8-$b$. Such high figures reflect the fact that for many polymorphs the vdW scheme allows one to find a qualitatively different geometry from the PBE one (e.g. internal voids and large-scale rings tend to be less symmetric and more puckered). This densification effect, which acts effectively as an internal negative pressure, results in an increasing enthalpic penalty with decreasing density (Fig.~\ref{fig_benchmark_4}, right panel). Overall, the more porous the polymorph, the larger the vdW contribution to the energy. This trend is supported by the RPA results presented in the same figure, and is also in line with studies of e.g. silica zeolites~\cite{hay_2015_1}. In other words, the energies of the predicted polymorphs are more impacted by vdW than the B$_2$O$_3$-I one, because of their lower density.

\begin{figure*}
\vskip -0.2cm
\includegraphics[height=8.3cm,width=18.2cm]{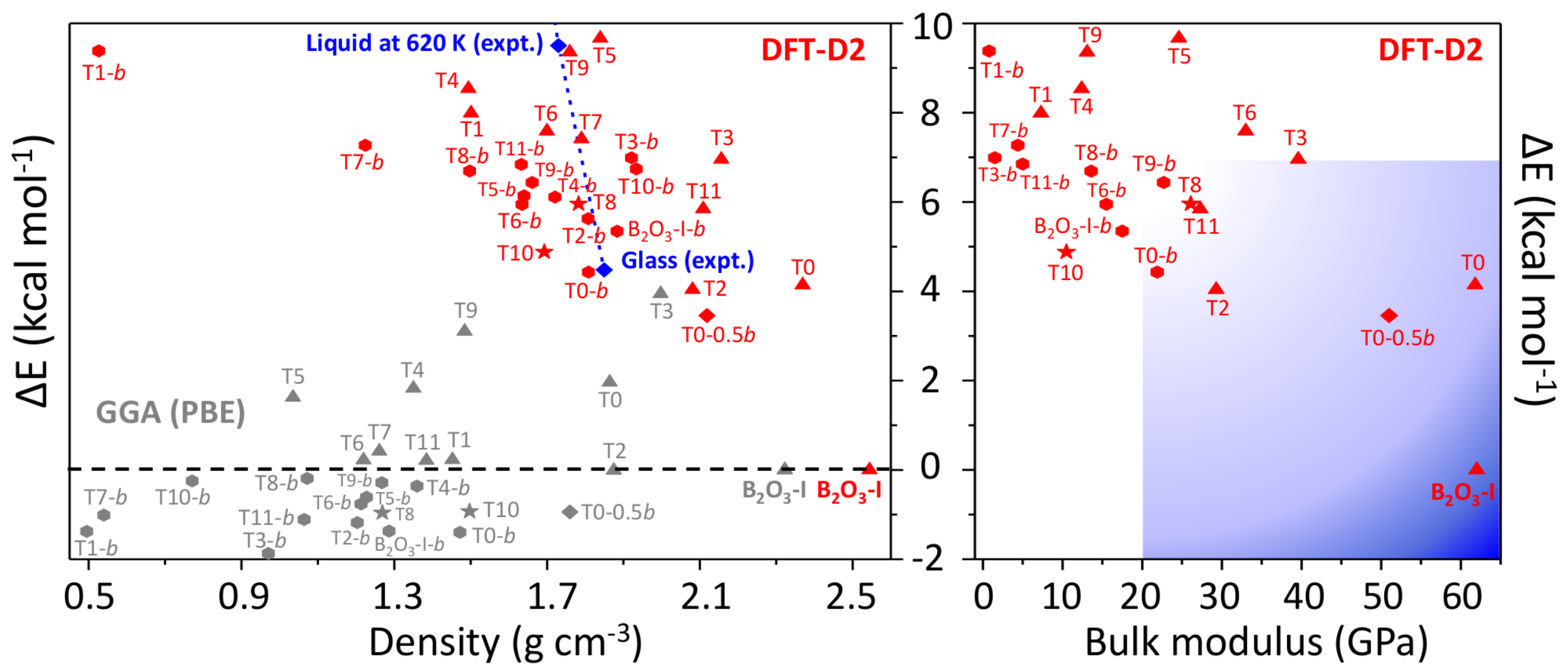}
\vskip -0.15cm
\caption{
Left: Energy, relative to ${\rm B_2O_3}$-I, and density for all polymorphs using DFT-D2 (red) and PBE (gray symbols). Symbols refer to the relative proportions of structural units (triangles:boroxol) in the polymorphs: $\triangle$ only BO$_3$ triangles (1:0), \protect\varhexagon \ only boroxol rings (0:1), \protect\emptystar mixed decoration (3:1), \protect$\lozenge$ \ mixed decoration (1:1). 
Experimental data (liquid and glass) are from Ref.~\protect\cite{shmi_66_1}.
Right: energy and bulk modulus using DFT-D2. The inner rectangle schematically represents the {\em feasibility} window.}
\label{fig_all_poly}
\end{figure*}

The energy diagram (Fig.~\ref{fig_all_poly}, left panel) resulting from the vdW inclusion provides several major outcomes. First, all predicted polymorphs energies fall above B$_2$O$_3$-I, while in the PBE picture~\cite{ferl_2012_1,clae_2013_1} many polymorphs are possible candidates for the ground state. Note that the overall correlation between metastability and density is now in line with the one observed for silica polymorphs~\cite{coud_2013_1}. Although the results are presented at 0 K, we checked that the picture obtained at 300 K is unchanged within 1 kcal/mol (by computing finite-temperature contributions to the vibrational free energy within the DFT-D2 scheme; see the SM).
A second important point is that, despite being metastable, most of the predicted polymorphs are still within a thermodynamically accessible energy range, estimated at $\sim$ 7 kcal/mol~\cite{upper_energy}. 
In particular, polymorphs such as T0-$0.5b$ and T0-$b$ could be amenable to synthesis. Remarkably, the layers that constitute these polymorphs are found experimentally in several chemically complex borates~\cite{wang_2007_1,liu_2009_2,daub_2015_1}. 

However, assessing the synthesizability of a given polymorph is a very challenging task, since the energy is not the sole ingredient at play. 
Following an earlier work on silica zeolites which associated experimentally observed structures with good mechanical properties~\cite{coud_2013_1}, we undertook a comprehensive characterization of the mechanical properties of the B$_2$O$_3$ polymorphs, including bulk ($B$), Young ($E$), and shear ($G$) moduli as well as linear compressibility and Poisson's ratio~\cite{hay_thesis}. 
In all the calculated moduli, and as illustrated in Fig.~\ref{fig_all_poly} (right panel) for the bulk modulus, there is a rather clear distinction between B$_2$O$_3$-I ($B \sim$ 60 GPa) and most of the other polymorphs which show much smaller values ($\sim$ 0-30 GPa). For illustrative purposes, the shaded-blue rectangle in the energy-versus-$B$ parameter space delimits the ranges of values in experimentally realized silica polymorphs. Noticeably, this area excludes most of the predicted B$_2$O$_3$ polymorphs, with a few noticeable exceptions including T0-$b$ and T0-0.5$b$.

Further, for most polymorphs, $E$ and $G$, which are directional quantities, show high anisotropies, implying that there is one direction much {\em weaker} than the others with respect to an applied stress. This is in line with the aforementioned {\em soft} direction in these structures.  
In silica zeolites~\cite{coud_2013_1}, proposals of {\em feasibility} criteria were derived from the lowest values of these mechanical quantities: strictly transferring these criteria~\cite{moduli_criteria} to B$_2$O$_3$ leaves only B$_2$O$_3$-I as a realizable structure with, however, a handful of additional candidates (T3, T11, B$_2$O$_3$-I-$b$, T0, and T0-0.5$b$) close to the criteria thresholds. Thus, many of the predicted structures suffer from mechanical stability issues and this may be one of the reasons why they have not been experimentally realized yet. 

Our findings have strong implications for understanding the B$_2$O$_3$ behavior at the glass transition. Indeed, the vast majority of the predicted polymorphs are clustered close to the liquid and glass states in the energy-density diagram (dashed-blue line in Fig.~\ref{fig_all_poly}, left panel). 
Since the glass structure is generally expected to resemble that of the underlying crystals, Fig.~\ref{fig_all_poly} provides a framework to understand the glass state and its apparent anomalous properties, such as its low density, its high fraction of rings, and its strong structural dissimilarity with B$_2$O$_3$-I. In addition, the existence of many competing polymorphs of similar energies, as observed here, has been shown to correlate with the glass forming ability; a situation reported for a large range of systems~\cite{peri_2016_1,good_75_1} and 
models~\cite{naum_2012_1,ronc_2017_1,russ_2018_1}.

These aspects, which combine the energetics with the mechanical properties, allow one to propose the following explanation for the B$_2$O$_3$ {\em crystallization anomaly}. As the low-density liquid is cooled and gets closer to the region of high polymorphic degeneracy, it hovers over a rugged energy landscape, i.e., dominated by many local minima of similar energies. These minima are associated to low-density and mechanically weak structures prone to collapse into an amorphous framework. Although there exists a ground state (B$_2$O$_3$-I) of higher density, the driving force (energy separation between the metastable states and B$_2$O$_3$-I) is small enough and the energy barriers (associated to the topological reconfiguration required to reach the B$_2$O$_3$-I density) are sufficiently high so that the system is kinetically trapped, in accordance with the very sluggish kinetics found in this system~\cite{zano_2017_1}. Applying pressure to the melt favors higher density and stiffer structures, and thus results in the B$_2$O$_3$-I crystallization.

\section{CONCLUSIONS}

We have revealed that B$_2$O$_3$ is a system remarkably sensitive to vdW interactions in both energies and structures. The importance of vdW in B$_2$O$_3$ is striking and unexpected, particularly when compared to SiO$_2$, a system commonly considered similar for which, however, only mild effects from vdW have been reported~\cite{hay_2015_1}. Thus, the set of polymorphs studied in this work constitutes a valuable test bed to develop new methods for the accurate treatment of electronic correlation. At the same time, the account of vdW allows one to retrieve a polymorphic picture which not only supports the generally admitted experimental knowledge (B$_2$O$_3$-I is the ground-state) but also brings the metastable polymorphs closer to the glass in the energy-density phase-space. 
From the simultaneous characterization of energies and mechanical properties, we provide here a semi-quantitative map of the {\em likeliness of synthesizability}. Not only is it in agreement with the experimental observations - the structural motif from the most robust predicted polymorph (T0-0.5$b$) has been found in synthetized borates - but also it provides a framework to explain the B$_2$O$_3$ intriguing anomalies. The glass has a low density and a high amount of boroxol rings because the supercooled liquid acquires those structural characteristics from the closest crystalline polymorphs. It, however, fails to crystallize in any of these because of both their energies degeneracy (which induces frustration) and their weak mechanical properties.


\vskip 0.3cm

We thank Harald Hillebrecht for communicating the results of Ref.~\cite{daub_2015_1} prior to their publication. This work was performed using HPC resources from GENCI-TGCC/CINES/IDRIS (Grants No. A0030906493, No. A0030801875, No. A0010906493, and No. A0010907625) and from PRACE (project 2016143322). Financial supports from the French National Research Agency program PIPOG ANR-17-CE30-000 and from the program Emergence-Ville de Paris are acknowledged.


\begin{thebibliography}{72}
\expandafter\ifx\csname natexlab\endcsname\relax\def\natexlab#1{#1}\fi
\expandafter\ifx\csname bibnamefont\endcsname\relax
  \def\bibnamefont#1{#1}\fi
\expandafter\ifx\csname bibfnamefont\endcsname\relax
  \def\bibfnamefont#1{#1}\fi
\expandafter\ifx\csname citenamefont\endcsname\relax
  \def\citenamefont#1{#1}\fi
\expandafter\ifx\csname url\endcsname\relax
  \def\url#1{\texttt{#1}}\fi
\expandafter\ifx\csname urlprefix\endcsname\relax\def\urlprefix{URL }\fi
\providecommand{\bibinfo}[2]{#2}
\providecommand{\eprint}[2][]{\url{#2}}

\bibitem[{\citenamefont{Ferlat}(2015)}]{ferl_2015_1}
\bibinfo{author}{\bibfnamefont{G.}~\bibnamefont{Ferlat}},
  \emph{\bibinfo{title}{Rings in Network Glasses: The B$_2$O$_3$ Case}}
  (\bibinfo{publisher}{Springer}, \bibinfo{address}{Switzerland},
  \bibinfo{year}{2015}), chap.~\bibinfo{chapter}{14}, p. \bibinfo{pages}{367}.

\bibitem[{\citenamefont{Wright}(2018)}]{wrig_2018_1}
\bibinfo{author}{\bibfnamefont{A.~C.} \bibnamefont{Wright}},
  \bibinfo{journal}{Phys. Chem. Glasses: Eur. J. Glass Sci. Technol. B}
  \textbf{\bibinfo{volume}{59}}, \bibinfo{pages}{65} (\bibinfo{year}{2018}).

\bibitem[{\citenamefont{Nicholas et~al.}(2004)\citenamefont{Nicholas,
  Sinogeikin, Kieffer, and Bass}}]{nich_2004_1}
\bibinfo{author}{\bibfnamefont{J.}~\bibnamefont{Nicholas}},
  \bibinfo{author}{\bibfnamefont{S.}~\bibnamefont{Sinogeikin}},
  \bibinfo{author}{\bibfnamefont{J.}~\bibnamefont{Kieffer}}, \bibnamefont{and}
  \bibinfo{author}{\bibfnamefont{J.}~\bibnamefont{Bass}},
  \bibinfo{journal}{Phys. Rev. Lett.} \textbf{\bibinfo{volume}{92}},
  \bibinfo{pages}{215701} (\bibinfo{year}{2004}).

\bibitem[{\citenamefont{Lee et~al.}(2005)\citenamefont{Lee, Mibe, Fei, Cody,
  and Mysen}}]{lee_2005_1}
\bibinfo{author}{\bibfnamefont{S.~K.} \bibnamefont{Lee}},
  \bibinfo{author}{\bibfnamefont{K.}~\bibnamefont{Mibe}},
  \bibinfo{author}{\bibfnamefont{Y.}~\bibnamefont{Fei}},
  \bibinfo{author}{\bibfnamefont{G.~D.} \bibnamefont{Cody}}, \bibnamefont{and}
  \bibinfo{author}{\bibfnamefont{B.~O.} \bibnamefont{Mysen}},
  \bibinfo{journal}{Phys. Rev. Lett.} \textbf{\bibinfo{volume}{94}},
  \bibinfo{pages}{165507} (\bibinfo{year}{2005}).

\bibitem[{\citenamefont{Trachenko et~al.}(2008)\citenamefont{Trachenko,
  Brazhkin, Ferlat, Dove, and Artacho}}]{trac_2008_1}
\bibinfo{author}{\bibfnamefont{K.}~\bibnamefont{Trachenko}},
  \bibinfo{author}{\bibfnamefont{V.~V.} \bibnamefont{Brazhkin}},
  \bibinfo{author}{\bibfnamefont{G.}~\bibnamefont{Ferlat}},
  \bibinfo{author}{\bibfnamefont{M.~T.} \bibnamefont{Dove}}, \bibnamefont{and}
  \bibinfo{author}{\bibfnamefont{E.}~\bibnamefont{Artacho}},
  \bibinfo{journal}{Phys. Rev. B} \textbf{\bibinfo{volume}{78}},
  \bibinfo{pages}{172102} (\bibinfo{year}{2008}).

\bibitem[{\citenamefont{Zeidler et~al.}(2014)\citenamefont{Zeidler, Wezka,
  Whittaker, Salmon, Baroni, Klotz, Fischer, Wilding, Bull, Tucker
  et~al.}}]{zeid_2014_1}
\bibinfo{author}{\bibfnamefont{A.}~\bibnamefont{Zeidler}},
  \bibinfo{author}{\bibfnamefont{K.}~\bibnamefont{Wezka}},
  \bibinfo{author}{\bibfnamefont{D.~A.~J.} \bibnamefont{Whittaker}},
  \bibinfo{author}{\bibfnamefont{S.~P.} \bibnamefont{Salmon}},
  \bibinfo{author}{\bibfnamefont{A.}~\bibnamefont{Baroni}},
  \bibinfo{author}{\bibfnamefont{S.}~\bibnamefont{Klotz}},
  \bibinfo{author}{\bibfnamefont{H.~E.} \bibnamefont{Fischer}},
  \bibinfo{author}{\bibfnamefont{M.~C.} \bibnamefont{Wilding}},
  \bibinfo{author}{\bibfnamefont{C.~L.} \bibnamefont{Bull}},
  \bibinfo{author}{\bibfnamefont{M.~G.} \bibnamefont{Tucker}},
  \bibnamefont{et~al.}, \bibinfo{journal}{Phys. Rev. B}
  \textbf{\bibinfo{volume}{90}}, \bibinfo{pages}{024206}
  (\bibinfo{year}{2014}).

\bibitem[{\citenamefont{Lee et~al.}(2018)\citenamefont{Lee, Kim, Chow, Xiao,
  Cheng, and Shen}}]{lee_2018_1}
\bibinfo{author}{\bibfnamefont{S.~K.} \bibnamefont{Lee}},
  \bibinfo{author}{\bibfnamefont{Y.-H.} \bibnamefont{Kim}},
  \bibinfo{author}{\bibfnamefont{P.}~\bibnamefont{Chow}},
  \bibinfo{author}{\bibfnamefont{Y.}~\bibnamefont{Xiao}},
  \bibinfo{author}{\bibfnamefont{J.}~\bibnamefont{Cheng}}, \bibnamefont{and}
  \bibinfo{author}{\bibfnamefont{G.}~\bibnamefont{Shen}},
  \bibinfo{journal}{Proc. Natl. Acad. Sci. USA} \textbf{\bibinfo{volume}{115}},
  \bibinfo{pages}{5855} (\bibinfo{year}{2018}).

\bibitem[{\citenamefont{Brazhkin et~al.}(2010)\citenamefont{Brazhkin, Farnan,
  Funakoshi, Kanzaki, Katayama, Lyapin, and Saitoh}}]{braz_2010_1}
\bibinfo{author}{\bibfnamefont{V.~V.} \bibnamefont{Brazhkin}},
  \bibinfo{author}{\bibfnamefont{I.}~\bibnamefont{Farnan}},
  \bibinfo{author}{\bibfnamefont{K.~I.} \bibnamefont{Funakoshi}},
  \bibinfo{author}{\bibfnamefont{M.}~\bibnamefont{Kanzaki}},
  \bibinfo{author}{\bibfnamefont{Y.}~\bibnamefont{Katayama}},
  \bibinfo{author}{\bibfnamefont{A.~G.} \bibnamefont{Lyapin}},
  \bibnamefont{and} \bibinfo{author}{\bibfnamefont{H.}~\bibnamefont{Saitoh}},
  \bibinfo{journal}{Phys. Rev. Lett.} \textbf{\bibinfo{volume}{105}},
  \bibinfo{pages}{115701} (\bibinfo{year}{2010}).

\bibitem[{\citenamefont{Alderman et~al.}(2015)\citenamefont{Alderman, Ferlat,
  Baroni, Salanne, Micoulaut, Benmore, Lin, Tamalonis, and
  Weber}}]{alde_2015_1}
\bibinfo{author}{\bibfnamefont{O.~L.~G.} \bibnamefont{Alderman}},
  \bibinfo{author}{\bibfnamefont{G.}~\bibnamefont{Ferlat}},
  \bibinfo{author}{\bibfnamefont{A.}~\bibnamefont{Baroni}},
  \bibinfo{author}{\bibfnamefont{M.}~\bibnamefont{Salanne}},
  \bibinfo{author}{\bibfnamefont{M.}~\bibnamefont{Micoulaut}},
  \bibinfo{author}{\bibfnamefont{C.~J.} \bibnamefont{Benmore}},
  \bibinfo{author}{\bibfnamefont{A.}~\bibnamefont{Lin}},
  \bibinfo{author}{\bibfnamefont{A.}~\bibnamefont{Tamalonis}},
  \bibnamefont{and} \bibinfo{author}{\bibfnamefont{J.~K.~R.}
  \bibnamefont{Weber}}, \bibinfo{journal}{J. Phys. Condens. Matter}
  \textbf{\bibinfo{volume}{27}}, \bibinfo{pages}{455104}
  (\bibinfo{year}{2015}).

\bibitem[{\citenamefont{Gurr et~al.}(1970)\citenamefont{Gurr, Montgomery,
  Knutson, and Gorres}}]{gurr_70_1}
\bibinfo{author}{\bibfnamefont{G.~E.} \bibnamefont{Gurr}},
  \bibinfo{author}{\bibfnamefont{P.~W.} \bibnamefont{Montgomery}},
  \bibinfo{author}{\bibfnamefont{C.~D.} \bibnamefont{Knutson}},
  \bibnamefont{and} \bibinfo{author}{\bibfnamefont{B.~T.}
  \bibnamefont{Gorres}}, \bibinfo{journal}{Acta Cristallogr.}
  \textbf{\bibinfo{volume}{B26}}, \bibinfo{pages}{906} (\bibinfo{year}{1970}).

\bibitem[{\citenamefont{Cole and Taylor}(1935)}]{cole_35_1}
\bibinfo{author}{\bibfnamefont{S.~S.} \bibnamefont{Cole}} \bibnamefont{and}
  \bibinfo{author}{\bibfnamefont{N.~W.} \bibnamefont{Taylor}},
  \bibinfo{journal}{J. Am. Ceram. Soc.} \textbf{\bibinfo{volume}{18}},
  \bibinfo{pages}{55} (\bibinfo{year}{1935}).

\bibitem[{\citenamefont{Piccione et~al.}(2000)\citenamefont{Piccione, Laberty,
  Yang, Camblor, Navrotsky, and Davis}}]{picc_2000_1}
\bibinfo{author}{\bibfnamefont{P.~M.} \bibnamefont{Piccione}},
  \bibinfo{author}{\bibfnamefont{C.}~\bibnamefont{Laberty}},
  \bibinfo{author}{\bibfnamefont{S.}~\bibnamefont{Yang}},
  \bibinfo{author}{\bibfnamefont{M.~A.} \bibnamefont{Camblor}},
  \bibinfo{author}{\bibfnamefont{A.}~\bibnamefont{Navrotsky}},
  \bibnamefont{and} \bibinfo{author}{\bibfnamefont{M.~E.} \bibnamefont{Davis}},
  \bibinfo{journal}{J. Phys. Chem. B} \textbf{\bibinfo{volume}{104}},
  \bibinfo{pages}{10001} (\bibinfo{year}{2000}).

\bibitem[{\citenamefont{Youngman et~al.}(1995)\citenamefont{Youngman, Haubrich,
  Zwanziger, Janicke, and Chmelka}}]{youn_95_1}
\bibinfo{author}{\bibfnamefont{R.~E.} \bibnamefont{Youngman}},
  \bibinfo{author}{\bibfnamefont{S.~T.} \bibnamefont{Haubrich}},
  \bibinfo{author}{\bibfnamefont{J.~W.} \bibnamefont{Zwanziger}},
  \bibinfo{author}{\bibfnamefont{M.~T.} \bibnamefont{Janicke}},
  \bibnamefont{and} \bibinfo{author}{\bibfnamefont{B.~F.}
  \bibnamefont{Chmelka}}, \bibinfo{journal}{Science}
  \textbf{\bibinfo{volume}{269}}, \bibinfo{pages}{1416} (\bibinfo{year}{1995}).

\bibitem[{\citenamefont{Umari and Pasquarello}(2005)}]{umar_2005_1}
\bibinfo{author}{\bibfnamefont{P.}~\bibnamefont{Umari}} \bibnamefont{and}
  \bibinfo{author}{\bibfnamefont{A.}~\bibnamefont{Pasquarello}},
  \bibinfo{journal}{Phys. Rev. Lett.} \textbf{\bibinfo{volume}{95}},
  \bibinfo{pages}{137401} (\bibinfo{year}{2005}).

\bibitem[{\citenamefont{Zanotto and Cassar}(2017)}]{zano_2017_1}
\bibinfo{author}{\bibfnamefont{E.~D.} \bibnamefont{Zanotto}} \bibnamefont{and}
  \bibinfo{author}{\bibfnamefont{D.~R.} \bibnamefont{Cassar}},
  \bibinfo{journal}{Sci. Rep.} \textbf{\bibinfo{volume}{7}},
  \bibinfo{pages}{43022} (\bibinfo{year}{2017}).

\bibitem[{\citenamefont{Ulhmann et~al.}(1967)\citenamefont{Ulhmann, Hays, and
  Turnbull}}]{uhlm_67_1}
\bibinfo{author}{\bibfnamefont{D.~R.} \bibnamefont{Ulhmann}},
  \bibinfo{author}{\bibfnamefont{J.~F.} \bibnamefont{Hays}}, \bibnamefont{and}
  \bibinfo{author}{\bibfnamefont{D.}~\bibnamefont{Turnbull}},
  \bibinfo{journal}{Phys. Chem. Glasses} \textbf{\bibinfo{volume}{8}},
  \bibinfo{pages}{1} (\bibinfo{year}{1967}).

\bibitem[{\citenamefont{Takada et~al.}(2003)\citenamefont{Takada, Catlow, and
  Price}}]{taka_2003_1}
\bibinfo{author}{\bibfnamefont{A.}~\bibnamefont{Takada}},
  \bibinfo{author}{\bibfnamefont{C.~R.~A.} \bibnamefont{Catlow}},
  \bibnamefont{and} \bibinfo{author}{\bibfnamefont{G.~D.} \bibnamefont{Price}},
  \bibinfo{journal}{Phys. Chem. Glasses} \textbf{\bibinfo{volume}{44}},
  \bibinfo{pages}{147} (\bibinfo{year}{2003}).

\bibitem[{\citenamefont{Huang et~al.}(2007)\citenamefont{Huang, Durandurdu, and
  Kieffer}}]{huan_2007_1}
\bibinfo{author}{\bibfnamefont{L.}~\bibnamefont{Huang}},
  \bibinfo{author}{\bibfnamefont{M.}~\bibnamefont{Durandurdu}},
  \bibnamefont{and} \bibinfo{author}{\bibfnamefont{J.}~\bibnamefont{Kieffer}},
  \bibinfo{journal}{J. Phys. Chem. C} \textbf{\bibinfo{volume}{111}},
  \bibinfo{pages}{13712} (\bibinfo{year}{2007}).

\bibitem[{\citenamefont{Ferlat et~al.}(2012)\citenamefont{Ferlat, Seitsonen,
  Lazzeri, and Mauri}}]{ferl_2012_1}
\bibinfo{author}{\bibfnamefont{G.}~\bibnamefont{Ferlat}},
  \bibinfo{author}{\bibfnamefont{A.~P.} \bibnamefont{Seitsonen}},
  \bibinfo{author}{\bibfnamefont{M.}~\bibnamefont{Lazzeri}}, \bibnamefont{and}
  \bibinfo{author}{\bibfnamefont{F.}~\bibnamefont{Mauri}},
  \bibinfo{journal}{Nature Mat.} \textbf{\bibinfo{volume}{11}},
  \bibinfo{pages}{925} (\bibinfo{year}{2012}).

\bibitem[{\citenamefont{Claeyssens et~al.}(2013)\citenamefont{Claeyssens, Hart,
  Norman, and Allan}}]{clae_2013_1}
\bibinfo{author}{\bibfnamefont{F.}~\bibnamefont{Claeyssens}},
  \bibinfo{author}{\bibfnamefont{J.~N.} \bibnamefont{Hart}},
  \bibinfo{author}{\bibfnamefont{N.~C.} \bibnamefont{Norman}},
  \bibnamefont{and} \bibinfo{author}{\bibfnamefont{N.~L.} \bibnamefont{Allan}},
  \bibinfo{journal}{Adv. Funct. Mater.} \textbf{\bibinfo{volume}{23}},
  \bibinfo{pages}{5887} (\bibinfo{year}{2013}).

\bibitem[{\citenamefont{Daub and Hillebrecht}(2015)}]{daub_2015_1}
\bibinfo{author}{\bibfnamefont{M.}~\bibnamefont{Daub}} \bibnamefont{and}
  \bibinfo{author}{\bibfnamefont{H.}~\bibnamefont{Hillebrecht}},
  \bibinfo{journal}{Eur. J. Inorg. Chem.} \textbf{\bibinfo{volume}{2015}},
  \bibinfo{pages}{4176} (\bibinfo{year}{2015}).

\bibitem[{\citenamefont{Sorella et~al.}(2007)\citenamefont{Sorella, Casula, and Rocca}}]{michele_benzene}
\bibinfo{author}{\bibfnamefont{S.}~\bibnamefont{Sorella}},
  \bibinfo{author}{\bibfnamefont{M.}~\bibnamefont{Casula}}, \bibnamefont{and}
  \bibinfo{author}{\bibfnamefont{D.}~\bibnamefont{Rocca}}, \bibinfo{journal}{J. Chem. Phys.} \textbf{\bibinfo{volume}{127}}, \bibinfo{pages}{014105}
  (\bibinfo{year}{2007}).

\bibitem[{\citenamefont{Umrigar et~al.}(2007)\citenamefont{Umrigar, Toulouse, Filippi, Sorella, and Hennig}}]{Umrigar2007}
\bibinfo{author}{\bibfnamefont{C.~J.} \bibnamefont{Umrigar}},
  \bibinfo{author}{\bibfnamefont{J.}~\bibnamefont{Toulouse}},
  \bibinfo{author}{\bibfnamefont{C.}~\bibnamefont{Filippi}},
  \bibinfo{author}{\bibfnamefont{S.}~\bibnamefont{Sorella}}, \bibnamefont{and}
  \bibinfo{author}{\bibfnamefont{R.~G.} \bibnamefont{Hennig}},
  \bibinfo{journal}{Phys. Rev. Lett.} \textbf{\bibinfo{volume}{98}},
  \bibinfo{pages}{110201}
  (\bibinfo{year}{2007}).

\bibitem[{\citenamefont{Barborini et~al.}(2012)\citenamefont{Barborini,
  Sorella, and Guidoni}}]{barborini2012structural}
\bibinfo{author}{\bibfnamefont{M.}~\bibnamefont{Barborini}},
  \bibinfo{author}{\bibfnamefont{S.}~\bibnamefont{Sorella}}, \bibnamefont{and}
  \bibinfo{author}{\bibfnamefont{L.}~\bibnamefont{Guidoni}},
  \bibinfo{journal}{J. Chem. Theory Comput.} \textbf{\bibinfo{volume}{8}},
  \bibinfo{pages}{1260} (\bibinfo{year}{2012}).

\bibitem[{\citenamefont{Casula et~al.}(2005)\citenamefont{Casula, Filippi, and
  Sorella}}]{michele_lrdmc}
\bibinfo{author}{\bibfnamefont{M.}~\bibnamefont{Casula}},
  \bibinfo{author}{\bibfnamefont{C.}~\bibnamefont{Filippi}}, \bibnamefont{and}
  \bibinfo{author}{\bibfnamefont{S.}~\bibnamefont{Sorella}},
  \bibinfo{journal}{Phys. Rev. Lett.} \textbf{\bibinfo{volume}{95}},
  \bibinfo{pages}{100201} (\bibinfo{year}{2005}).

\bibitem[{\citenamefont{Casula et~al.}(2010)\citenamefont{Casula, Moroni,
  Sorella, and Filippi}}]{filippi_lrdmc}
\bibinfo{author}{\bibfnamefont{M.}~\bibnamefont{Casula}},
  \bibinfo{author}{\bibfnamefont{S.}~\bibnamefont{Moroni}},
  \bibinfo{author}{\bibfnamefont{S.}~\bibnamefont{Sorella}}, \bibnamefont{and}
  \bibinfo{author}{\bibfnamefont{C.}~\bibnamefont{Filippi}},
  \bibinfo{journal}{~J. Chem. Phys.} \textbf{\bibinfo{volume}{132}},
  \bibinfo{pages}{154113} (\bibinfo{year}{2010}).

\bibitem[{\citenamefont{Dagrada et~al.}(2016)\citenamefont{Dagrada, Karakuzu,
  Vildosola, Casula, and Sorella}}]{dagrada2016exact}
\bibinfo{author}{\bibfnamefont{M.}~\bibnamefont{Dagrada}},
  \bibinfo{author}{\bibfnamefont{S.}~\bibnamefont{Karakuzu}},
  \bibinfo{author}{\bibfnamefont{V.~L.} \bibnamefont{Vildosola}},
  \bibinfo{author}{\bibfnamefont{M.}~\bibnamefont{Casula}}, \bibnamefont{and}
  \bibinfo{author}{\bibfnamefont{S.}~\bibnamefont{Sorella}},
  \bibinfo{journal}{Phys. Rev. B} \textbf{\bibinfo{volume}{94}},
  \bibinfo{pages}{245108} (\bibinfo{year}{2016}).

\bibitem[{\citenamefont{Sorella}()}]{sorella_rvb}
\bibinfo{author}{\bibfnamefont{S.}~\bibnamefont{Sorella}},
  \emph{\bibinfo{title}{Turborvb, quantum monte carlo software for electronic
  structure calculations}},
  \bibinfo{howpublished}{http://people.sissa.it/~sorella/web/}.


\bibitem[{\citenamefont{Langreth and Perdew}(1975)}]{lang_75_1}
\bibinfo{author}{\bibfnamefont{D.~C.} \bibnamefont{Langreth}} \bibnamefont{and}
  \bibinfo{author}{\bibfnamefont{J.~P.} \bibnamefont{Perdew}},
  \bibinfo{journal}{Solid State Comm.} \textbf{\bibinfo{volume}{17}},
  \bibinfo{pages}{1425} (\bibinfo{year}{1975}).

\bibitem[{\citenamefont{Gunnarsson and Lundqvist}(1976)}]{gunn_76_1}
\bibinfo{author}{\bibfnamefont{O.}~\bibnamefont{Gunnarsson}} \bibnamefont{and}
  \bibinfo{author}{\bibfnamefont{B.~I.} \bibnamefont{Lundqvist}},
  \bibinfo{journal}{Phys. Rev. B} \textbf{\bibinfo{volume}{13}},
  \bibinfo{pages}{4274} (\bibinfo{year}{1976}).

\bibitem[{\citenamefont{Marini et~al.}(2006)\citenamefont{Marini,
  Garc\'{\i}a-Gonz\'alez, and Rubio}}]{BN1}
\bibinfo{author}{\bibfnamefont{A.}~\bibnamefont{Marini}},
  \bibinfo{author}{\bibfnamefont{P.}~\bibnamefont{Garc\'{\i}a-Gonz\'alez}},
  \bibnamefont{and} \bibinfo{author}{\bibfnamefont{A.}~\bibnamefont{Rubio}},
  \bibinfo{journal}{Phys. Rev. Lett.} \textbf{\bibinfo{volume}{96}},
  \bibinfo{pages}{136404} (\bibinfo{year}{2006}).

\bibitem[{\citenamefont{Leb\`egue et~al.}(2010)\citenamefont{Leb\`egue, Harl,
  Gould, \'Angy\'an, Kresse, and Dobson}}]{GRAPHRPA}
\bibinfo{author}{\bibfnamefont{S.}~\bibnamefont{Leb\`egue}},
  \bibinfo{author}{\bibfnamefont{J.}~\bibnamefont{Harl}},
  \bibinfo{author}{\bibfnamefont{T.}~\bibnamefont{Gould}},
  \bibinfo{author}{\bibfnamefont{J.~G.} \bibnamefont{\'Angy\'an}},
  \bibinfo{author}{\bibfnamefont{G.}~\bibnamefont{Kresse}}, \bibnamefont{and}
  \bibinfo{author}{\bibfnamefont{J.~F.} \bibnamefont{Dobson}},
  \bibinfo{journal}{Phys. Rev. Lett.} \textbf{\bibinfo{volume}{105}},
  \bibinfo{pages}{196401} (\bibinfo{year}{2010}).

\bibitem[{\citenamefont{Kresse and Furthm\"uller}(1996)}]{kres_96_1}
\bibinfo{author}{\bibfnamefont{G.}~\bibnamefont{Kresse}} \bibnamefont{and}
  \bibinfo{author}{\bibfnamefont{J.}~\bibnamefont{Furthm\"uller}},
  \bibinfo{journal}{Phys. Rev. B} \textbf{\bibinfo{volume}{54}},
  \bibinfo{pages}{11169} (\bibinfo{year}{1996}).

\bibitem[{\citenamefont{Kresse and Joubert}(1999)}]{kres_99_1}
\bibinfo{author}{\bibfnamefont{G.}~\bibnamefont{Kresse}} \bibnamefont{and}
  \bibinfo{author}{\bibfnamefont{D.}~\bibnamefont{Joubert}},
  \bibinfo{journal}{Phys. Rev. B} \textbf{\bibinfo{volume}{59}},
  \bibinfo{pages}{1758} (\bibinfo{year}{1999}).

\bibitem[{\citenamefont{Grimme}(2006)}]{grim_2006_1}
\bibinfo{author}{\bibfnamefont{S.}~\bibnamefont{Grimme}}, \bibinfo{journal}{J.
  Comput. Chem.} \textbf{\bibinfo{volume}{27}}, \bibinfo{pages}{1787}
  (\bibinfo{year}{2006}).

\bibitem[{\citenamefont{Tkatchenko and Scheffler}(2009)}]{tkat_2009_1}
\bibinfo{author}{\bibfnamefont{A.}~\bibnamefont{Tkatchenko}} \bibnamefont{and}
  \bibinfo{author}{\bibfnamefont{M.}~\bibnamefont{Scheffler}},
  \bibinfo{journal}{Phys. Rev. Lett.} \textbf{\bibinfo{volume}{102}},
  \bibinfo{pages}{073005} (\bibinfo{year}{2009}).

\bibitem[{\citenamefont{Lee et~al.}(2010)\citenamefont{Lee, Murray, Kong,
  Lundqvist, and Langreth}}]{lee_2010_1}
\bibinfo{author}{\bibfnamefont{K.}~\bibnamefont{Lee}},
  \bibinfo{author}{\bibfnamefont{E.~D.} \bibnamefont{Murray}},
  \bibinfo{author}{\bibfnamefont{L.}~\bibnamefont{Kong}},
  \bibinfo{author}{\bibfnamefont{B.~I.} \bibnamefont{Lundqvist}},
  \bibnamefont{and} \bibinfo{author}{\bibfnamefont{D.~C.}
  \bibnamefont{Langreth}}, \bibinfo{journal}{Phys. Rev. B}
  \textbf{\bibinfo{volume}{82}}, \bibinfo{pages}{081101(R)}
  (\bibinfo{year}{2010}).

\bibitem[{\citenamefont{Thonhauser et~al.}(2015)\citenamefont{Thonhauser,
  Zuluaga, Arter, Berland, Schr\"oder, and Hyldgaard}}]{thon_2015_1}
\bibinfo{author}{\bibfnamefont{T.}~\bibnamefont{Thonhauser}},
  \bibinfo{author}{\bibfnamefont{S.}~\bibnamefont{Zuluaga}},
  \bibinfo{author}{\bibfnamefont{C.~A.} \bibnamefont{Arter}},
  \bibinfo{author}{\bibfnamefont{K.}~\bibnamefont{Berland}},
  \bibinfo{author}{\bibfnamefont{E.}~\bibnamefont{Schr\"oder}},
  \bibnamefont{and}
  \bibinfo{author}{\bibfnamefont{P.}~\bibnamefont{Hyldgaard}},
  \bibinfo{journal}{Phys. Rev. Lett.} \textbf{\bibinfo{volume}{115}},
  \bibinfo{pages}{136402} (\bibinfo{year}{2015}).

\bibitem[{\citenamefont{Perdew et~al.}(1996)\citenamefont{Perdew, Burke, and
  Ernzerhof}}]{perd_96_1}
\bibinfo{author}{\bibfnamefont{J.~P.} \bibnamefont{Perdew}},
  \bibinfo{author}{\bibfnamefont{K.}~\bibnamefont{Burke}}, \bibnamefont{and}
  \bibinfo{author}{\bibfnamefont{M.}~\bibnamefont{Ernzerhof}},
  \bibinfo{journal}{Phys. Rev. Lett.} \textbf{\bibinfo{volume}{77}},
  \bibinfo{pages}{3865} (\bibinfo{year}{1996}).

\bibitem[{\citenamefont{Clark et~al.}(2005)\citenamefont{Clark, Segall,
  Pickard, Hasnip, Probert, Refson, and Payne}}]{clar_2005_1}
\bibinfo{author}{\bibfnamefont{S.~J.} \bibnamefont{Clark}},
  \bibinfo{author}{\bibfnamefont{M.~D.} \bibnamefont{Segall}},
  \bibinfo{author}{\bibfnamefont{C.~J.} \bibnamefont{Pickard}},
  \bibinfo{author}{\bibfnamefont{P.~J.} \bibnamefont{Hasnip}},
  \bibinfo{author}{\bibfnamefont{M.~J.} \bibnamefont{Probert}},
  \bibinfo{author}{\bibfnamefont{K.}~\bibnamefont{Refson}}, \bibnamefont{and}
  \bibinfo{author}{\bibfnamefont{M.~C.} \bibnamefont{Payne}},
  \bibinfo{journal}{Z. Kristallogr.} \textbf{\bibinfo{volume}{220}},
  \bibinfo{pages}{567} (\bibinfo{year}{2005}).

\bibitem[{\citenamefont{Giannozzi et~al.}(2009)\citenamefont{Giannozzi, Baroni,
  Bonini, Calandra, Car, Cavazzoni, Ceresoli, Chiarotti, Cococcioni, Dabo
  et~al.}}]{QE-2009}
\bibinfo{author}{\bibfnamefont{P.}~\bibnamefont{Giannozzi}},
  \bibinfo{author}{\bibfnamefont{S.}~\bibnamefont{Baroni}},
  \bibinfo{author}{\bibfnamefont{N.}~\bibnamefont{Bonini}},
  \bibinfo{author}{\bibfnamefont{M.}~\bibnamefont{Calandra}},
  \bibinfo{author}{\bibfnamefont{R.}~\bibnamefont{Car}},
  \bibinfo{author}{\bibfnamefont{C.}~\bibnamefont{Cavazzoni}},
  \bibinfo{author}{\bibfnamefont{D.}~\bibnamefont{Ceresoli}},
  \bibinfo{author}{\bibfnamefont{G.~L.} \bibnamefont{Chiarotti}},
  \bibinfo{author}{\bibfnamefont{M.}~\bibnamefont{Cococcioni}},
  \bibinfo{author}{\bibfnamefont{I.}~\bibnamefont{Dabo}}, \bibnamefont{et~al.},
  \bibinfo{journal}{J. Phys. Condens. Matter} \textbf{\bibinfo{volume}{21}},
  \bibinfo{pages}{395502} (\bibinfo{year}{2009}),
  \urlprefix\url{http://www.quantum-espresso.org}.

\bibitem[{\citenamefont{Coudert}(2013)}]{coud_2013_1}
\bibinfo{author}{\bibfnamefont{F.-X.} \bibnamefont{Coudert}},
  \bibinfo{journal}{Phys. Chem. Chem. Phys.} \textbf{\bibinfo{volume}{15}},
  \bibinfo{pages}{16012} (\bibinfo{year}{2013}).

\bibitem[{\citenamefont{Hay}(2016)}]{hay_thesis}
\bibinfo{author}{\bibfnamefont{H.}~\bibnamefont{Hay}}, Ph.D. thesis,
  \bibinfo{school}{Universit\'e Pierre \& Marie Curie},
  \bibinfo{address}{Paris} (\bibinfo{year}{2016}),
  \urlprefix\url{https://tel.archives-ouvertes.fr/tel-0147013a1}.

\bibitem[{\citenamefont{Haas et~al.}(2009)\citenamefont{Haas, Tran, and
  Blaha}}]{haas_2009_1}
\bibinfo{author}{\bibfnamefont{P.}~\bibnamefont{Haas}},
  \bibinfo{author}{\bibfnamefont{F.}~\bibnamefont{Tran}}, \bibnamefont{and}
  \bibinfo{author}{\bibfnamefont{P.}~\bibnamefont{Blaha}},
  \bibinfo{journal}{Phys. Rev. B} \textbf{\bibinfo{volume}{79}},
  \bibinfo{pages}{085104} (\bibinfo{year}{2009}).

\bibitem[{\citenamefont{Hay et~al.}(2015)\citenamefont{Hay, Ferlat, Casula,
  Seitsonen, and Mauri}}]{hay_2015_1}
\bibinfo{author}{\bibfnamefont{H.}~\bibnamefont{Hay}},
  \bibinfo{author}{\bibfnamefont{G.}~\bibnamefont{Ferlat}},
  \bibinfo{author}{\bibfnamefont{M.}~\bibnamefont{Casula}},
  \bibinfo{author}{\bibfnamefont{A.~P.} \bibnamefont{Seitsonen}},
  \bibnamefont{and} \bibinfo{author}{\bibfnamefont{F.}~\bibnamefont{Mauri}},
  \bibinfo{journal}{Phys. Rev. B} \textbf{\bibinfo{volume}{92}},
  \bibinfo{pages}{144111} (\bibinfo{year}{2015}).

\bibitem[{\citenamefont{Sun et~al.}(2016)\citenamefont{Sun, Dacek, Ong,
  Hautier, Jain, Richards, Gamst, Persson, and Ceder}}]{sun_2016_1}
\bibinfo{author}{\bibfnamefont{W.}~\bibnamefont{Sun}},
  \bibinfo{author}{\bibfnamefont{S.~T.} \bibnamefont{Dacek}},
  \bibinfo{author}{\bibfnamefont{S.~P.} \bibnamefont{Ong}},
  \bibinfo{author}{\bibfnamefont{G.}~\bibnamefont{Hautier}},
  \bibinfo{author}{\bibfnamefont{A.}~\bibnamefont{Jain}},
  \bibinfo{author}{\bibfnamefont{W.~D.} \bibnamefont{Richards}},
  \bibinfo{author}{\bibfnamefont{A.~C.} \bibnamefont{Gamst}},
  \bibinfo{author}{\bibfnamefont{K.~A.} \bibnamefont{Persson}},
  \bibnamefont{and} \bibinfo{author}{\bibfnamefont{G.}~\bibnamefont{Ceder}},
  \bibinfo{journal}{Sci. Adv.} \textbf{\bibinfo{volume}{2}},
  \bibinfo{pages}{e1600225} (\bibinfo{year}{2016}).

\bibitem[{BM_()}]{BM_D2}
\bibinfo{note}{The calculation of the mechanical moduli
  (Fig.~\ref{fig_all_poly}) is readily available in the CRYSTAL14
  code~\cite{dove_2014_1} for D2 only (among vdW schemes).}

\bibitem[{upp()}]{upper_energy}
\bibinfo{note}{For binary oxides, the range of observed polymorphs (as defined
  by the 90$^{th}$ percentile of a statistical analysis~\cite{sun_2016_1}) is
  94 meV/atom, i.e. 10.8 kcal/(mol B$_2$O$_3$). Taking silica as a close parent
  system, the highest energy above quartz (among SiO$_2$ polymorphs for which
  calorimetric data are available) is 6.88 kcal/(mol 2SiO$_2$) (ISV
  zeolite)~\cite{picc_2000_1}.}

\bibitem[{\citenamefont{Wang et~al.}(2007)\citenamefont{Wang, Guo, Chen, Xu,
  Zhou, Wu, and Huang}}]{wang_2007_1}
\bibinfo{author}{\bibfnamefont{M.-S.} \bibnamefont{Wang}},
  \bibinfo{author}{\bibfnamefont{G.-C.} \bibnamefont{Guo}},
  \bibinfo{author}{\bibfnamefont{W.-T.} \bibnamefont{Chen}},
  \bibinfo{author}{\bibfnamefont{G.}~\bibnamefont{Xu}},
  \bibinfo{author}{\bibfnamefont{W.-W.} \bibnamefont{Zhou}},
  \bibinfo{author}{\bibfnamefont{K.-J.} \bibnamefont{Wu}}, \bibnamefont{and}
  \bibinfo{author}{\bibfnamefont{J.-S.} \bibnamefont{Huang}},
  \bibinfo{journal}{Angew. Chem. Int. Ed.} \textbf{\bibinfo{volume}{46}},
  \bibinfo{pages}{3909} (\bibinfo{year}{2007}).

\bibitem[{\citenamefont{Liu et~al.}(2009)\citenamefont{Liu, Zhou, Yao, Ji,
  Zhang, Ji, and An}}]{liu_2009_2}
\bibinfo{author}{\bibfnamefont{M.-C.} \bibnamefont{Liu}},
  \bibinfo{author}{\bibfnamefont{P.}~\bibnamefont{Zhou}},
  \bibinfo{author}{\bibfnamefont{H.~G.} \bibnamefont{Yao}},
  \bibinfo{author}{\bibfnamefont{S.-H.} \bibnamefont{Ji}},
  \bibinfo{author}{\bibfnamefont{R.-C.} \bibnamefont{Zhang}},
  \bibinfo{author}{\bibfnamefont{M.}~\bibnamefont{Ji}}, \bibnamefont{and}
  \bibinfo{author}{\bibfnamefont{Y.-L.} \bibnamefont{An}},
  \bibinfo{journal}{Eur. J. Inorg. Chem.} \textbf{\bibinfo{volume}{31}},
  \bibinfo{pages}{4622} (\bibinfo{year}{2009}).

\bibitem[{\citenamefont{Shmidt}(1966)}]{shmi_66_1}
\bibinfo{author}{\bibfnamefont{N.~E.} \bibnamefont{Shmidt}},
  \bibinfo{journal}{Russ. J. Inorg. Chem.} \textbf{\bibinfo{volume}{11}},
  \bibinfo{pages}{241} (\bibinfo{year}{1966}).

\bibitem[{mod()}]{moduli_criteria}
\bibinfo{note}{Minimal value of the Young modulus $E_{min} \ge$ 20 GPa and
  elastic anisotropy $\eta \le $ 5, where $\eta$ is $\displaystyle \max \left (
  \frac{E_{max}}{E_{min}}, \frac{G_{max}}{G_{min}}\right )$. However, the
  criterion on anisotropy likely does not apply to layered structures (T0,
  T0-0.5$b$ and T0-$b$).}

\bibitem[{\citenamefont{Perim et~al.}(2016)\citenamefont{Perim, Lee, Liu,
  Toher, Gong, Li, Simmons, Levy, Vlassak, Schroers et~al.}}]{peri_2016_1}
\bibinfo{author}{\bibfnamefont{E.}~\bibnamefont{Perim}},
  \bibinfo{author}{\bibfnamefont{D.}~\bibnamefont{Lee}},
  \bibinfo{author}{\bibfnamefont{Y.}~\bibnamefont{Liu}},
  \bibinfo{author}{\bibfnamefont{C.}~\bibnamefont{Toher}},
  \bibinfo{author}{\bibfnamefont{P.}~\bibnamefont{Gong}},
  \bibinfo{author}{\bibfnamefont{Y.}~\bibnamefont{Li}},
  \bibinfo{author}{\bibfnamefont{W.~N.} \bibnamefont{Simmons}},
  \bibinfo{author}{\bibfnamefont{O.}~\bibnamefont{Levy}},
  \bibinfo{author}{\bibfnamefont{J.~J.} \bibnamefont{Vlassak}},
  \bibinfo{author}{\bibfnamefont{J.}~\bibnamefont{Schroers}},
  \bibnamefont{et~al.}, \bibinfo{journal}{Nat. Commun.}
  \textbf{\bibinfo{volume}{7}}, \bibinfo{pages}{12315} (\bibinfo{year}{2016}).

\bibitem[{\citenamefont{Goodman}(1975)}]{good_75_1}
\bibinfo{author}{\bibfnamefont{C.~H.~L.} \bibnamefont{Goodman}},
  \bibinfo{journal}{Nature} \textbf{\bibinfo{volume}{257}},
  \bibinfo{pages}{370} (\bibinfo{year}{1975}).

\bibitem[{\citenamefont{Naumis}(2012)}]{naum_2012_1}
\bibinfo{author}{\bibfnamefont{G.~G.} \bibnamefont{Naumis}},
  \bibinfo{journal}{Phys. Rev. E} \textbf{\bibinfo{volume}{85}},
  \bibinfo{pages}{061505} (\bibinfo{year}{2012}).

\bibitem[{\citenamefont{Ronceray and Harrowell}(2017)}]{ronc_2017_1}
\bibinfo{author}{\bibfnamefont{P.}~\bibnamefont{Ronceray}} \bibnamefont{and}
  \bibinfo{author}{\bibfnamefont{P.}~\bibnamefont{Harrowell}},
  \bibinfo{journal}{Phys. Rev. E} \textbf{\bibinfo{volume}{96}},
  \bibinfo{pages}{042602} (\bibinfo{year}{2017}).

\bibitem[{\citenamefont{Russo et~al.}(2018)\citenamefont{Russo, Romano, and
  Tanaka}}]{russ_2018_1}
\bibinfo{author}{\bibfnamefont{J.}~\bibnamefont{Russo}},
  \bibinfo{author}{\bibfnamefont{F.}~\bibnamefont{Romano}}, \bibnamefont{and}
  \bibinfo{author}{\bibfnamefont{H.}~\bibnamefont{Tanaka}},
  \bibinfo{journal}{Phys. Rev. X} \textbf{\bibinfo{volume}{8}},
  \bibinfo{pages}{021040} (\bibinfo{year}{2018}).

\bibitem[{\citenamefont{Dovesi et~al.}(2014)\citenamefont{Dovesi, Orlando,
  Erba, Zicovich-Wilson, Civalleri, Casassa, Maschio, Ferrabone, De~La~Pierre,
  D'Arco et~al.}}]{dove_2014_1}
\bibinfo{author}{\bibfnamefont{R.}~\bibnamefont{Dovesi}},
  \bibinfo{author}{\bibfnamefont{R.}~\bibnamefont{Orlando}},
  \bibinfo{author}{\bibfnamefont{A.}~\bibnamefont{Erba}},
  \bibinfo{author}{\bibfnamefont{C.~M.} \bibnamefont{Zicovich-Wilson}},
  \bibinfo{author}{\bibfnamefont{B.}~\bibnamefont{Civalleri}},
  \bibinfo{author}{\bibfnamefont{S.}~\bibnamefont{Casassa}},
  \bibinfo{author}{\bibfnamefont{L.}~\bibnamefont{Maschio}},
  \bibinfo{author}{\bibfnamefont{M.}~\bibnamefont{Ferrabone}},
  \bibinfo{author}{\bibfnamefont{M.}~\bibnamefont{De~La~Pierre}},
  \bibinfo{author}{\bibfnamefont{P.}~\bibnamefont{D'Arco}},
  \bibnamefont{et~al.}, \bibinfo{journal}{Int. J. Quantum Chem.}
  \textbf{\bibinfo{volume}{114}}, \bibinfo{pages}{1287} (\bibinfo{year}{2014}).
 
 
  
\bibitem[{\citenamefont{Winkler et~al.}(2001)\citenamefont{Winkler, Pickard,
  Milman, and Thimm}}]{wink_2001_1}
\bibinfo{author}{\bibfnamefont{B.}~\bibnamefont{Winkler}},
  \bibinfo{author}{\bibfnamefont{C.~J.} \bibnamefont{Pickard}},
  \bibinfo{author}{\bibfnamefont{V.}~\bibnamefont{Milman}}, \bibnamefont{and}
  \bibinfo{author}{\bibfnamefont{G.}~\bibnamefont{Thimm}},
  \bibinfo{journal}{Chem. Phys. Lett.} \textbf{\bibinfo{volume}{337}},
  \bibinfo{pages}{36} (\bibinfo{year}{2001}).
  
\bibitem[{\citenamefont{Vanderbilt}(1990)}]{vand_90_1}
\bibinfo{author}{\bibfnamefont{D.}~\bibnamefont{Vanderbilt}},
  \bibinfo{journal}{Phys. Rev. B} \textbf{\bibinfo{volume}{41}},
  \bibinfo{pages}{7892} (\bibinfo{year}{1990}).  

\bibitem[{\citenamefont{Richet et~al.}(1982)\citenamefont{Richet, Bottinga,
  Denielou, Petitet, and Tequi}}]{rich_82_1}
\bibinfo{author}{\bibfnamefont{P.}~\bibnamefont{Richet}},
  \bibinfo{author}{\bibfnamefont{Y.}~\bibnamefont{Bottinga}},
  \bibinfo{author}{\bibfnamefont{L.}~\bibnamefont{Denielou}},
  \bibinfo{author}{\bibfnamefont{J.}~\bibnamefont{Petitet}}, \bibnamefont{and}
  \bibinfo{author}{\bibfnamefont{C.}~\bibnamefont{Tequi}},
  \bibinfo{journal}{Geochim. Cosmochim. Acta} \textbf{\bibinfo{volume}{46}},
  \bibinfo{pages}{2639} (\bibinfo{year}{1982}).

\bibitem[{\citenamefont{Mouhat and Coudert}(2014)}]{mouh_2014_1}
\bibinfo{author}{\bibfnamefont{F.}~\bibnamefont{Mouhat}} \bibnamefont{and}
  \bibinfo{author}{\bibfnamefont{F.-X.} \bibnamefont{Coudert}},
  \bibinfo{journal}{Phys. Rev. B} \textbf{\bibinfo{volume}{90}},
  \bibinfo{pages}{224104} (\bibinfo{year}{2014}).

\bibitem[{\citenamefont{Gaillac et~al.}(2016)\citenamefont{Gaillac, Pullumbi,
  and Coudert}}]{gail_2016_1}
\bibinfo{author}{\bibfnamefont{R.}~\bibnamefont{Gaillac}},
  \bibinfo{author}{\bibfnamefont{P.}~\bibnamefont{Pullumbi}}, \bibnamefont{and}
  \bibinfo{author}{\bibfnamefont{F.-X.} \bibnamefont{Coudert}},
  \bibinfo{journal}{J. Phys. Condens. Matter} \textbf{\bibinfo{volume}{28}},
  \bibinfo{pages}{275201} (\bibinfo{year}{2016}).

\bibitem[{\citenamefont{Burkatzki et~al.}(2007)\citenamefont{Burkatzki,
  Filippi, and Dolg}}]{filippi_pseudo}
\bibinfo{author}{\bibfnamefont{M.}~\bibnamefont{Burkatzki}},
  \bibinfo{author}{\bibfnamefont{C.}~\bibnamefont{Filippi}}, \bibnamefont{and}
  \bibinfo{author}{\bibfnamefont{M.}~\bibnamefont{Dolg}}, \bibinfo{journal}{J.
  Chem. Phys.} \textbf{\bibinfo{volume}{126}}, \bibinfo{pages}{234105}
  (\bibinfo{year}{2007}).

\bibitem[{\citenamefont{Casula et~al.}(2004)\citenamefont{Casula, Attaccalite,
  and Sorella}}]{Casula2004}
\bibinfo{author}{\bibfnamefont{M.}~\bibnamefont{Casula}},
  \bibinfo{author}{\bibfnamefont{C.}~\bibnamefont{Attaccalite}},
  \bibnamefont{and} \bibinfo{author}{\bibfnamefont{S.}~\bibnamefont{Sorella}},
  \bibinfo{journal}{J. Chem. Phys.} \textbf{\bibinfo{volume}{121}},
  \bibinfo{pages}{7110} (\bibinfo{year}{2004}).

\bibitem[{\citenamefont{Marchi et~al.}(2009)\citenamefont{Marchi, Azadi,
  Casula, and Sorella}}]{michele_agp2}
\bibinfo{author}{\bibfnamefont{M.}~\bibnamefont{Marchi}},
  \bibinfo{author}{\bibfnamefont{S.}~\bibnamefont{Azadi}},
  \bibinfo{author}{\bibfnamefont{M.}~\bibnamefont{Casula}}, \bibnamefont{and}
  \bibinfo{author}{\bibfnamefont{S.}~\bibnamefont{Sorella}},
  \bibinfo{journal}{~J. Chem. Phys.} \textbf{\bibinfo{volume}{131}},
  \bibinfo{pages}{154116} (\bibinfo{year}{2009}).

\bibitem[{\citenamefont{Devaux et~al.}(2015)\citenamefont{Devaux, Casula,
  Decremps, and Sorella}}]{devaux2015electronic}
\bibinfo{author}{\bibfnamefont{N.}~\bibnamefont{Devaux}},
  \bibinfo{author}{\bibfnamefont{M.}~\bibnamefont{Casula}},
  \bibinfo{author}{\bibfnamefont{F.}~\bibnamefont{Decremps}}, \bibnamefont{and}
  \bibinfo{author}{\bibfnamefont{S.}~\bibnamefont{Sorella}},
  \bibinfo{journal}{Phys. Rev. B} \textbf{\bibinfo{volume}{91}},
  \bibinfo{pages}{081101} (\bibinfo{year}{2015}).

\bibitem[{\citenamefont{Calandra~Buonaura and Sorella}(1998)}]{Calandra1998}
\bibinfo{author}{\bibfnamefont{M.}~\bibnamefont{Calandra~Buonaura}}
  \bibnamefont{and} \bibinfo{author}{\bibfnamefont{S.}~\bibnamefont{Sorella}},
  \bibinfo{journal}{Phys. Rev. B} \textbf{\bibinfo{volume}{57}},
  \bibinfo{pages}{11446} (\bibinfo{year}{1998}).

\bibitem[{\citenamefont{Assaraf and Caffarel}(2003)}]{estim1}
\bibinfo{author}{\bibfnamefont{R.}~\bibnamefont{Assaraf}} \bibnamefont{and}
  \bibinfo{author}{\bibfnamefont{M.}~\bibnamefont{Caffarel}},
  \bibinfo{journal}{J. Chem. Phys.} \textbf{\bibinfo{volume}{119}},
  \bibinfo{pages}{10536} (\bibinfo{year}{2003}).

\bibitem[{\citenamefont{Filippi and Umrigar}(2000)}]{umrigar_warp}
\bibinfo{author}{\bibfnamefont{C.}~\bibnamefont{Filippi}} \bibnamefont{and}
  \bibinfo{author}{\bibfnamefont{C.~J.} \bibnamefont{Umrigar}},
  \bibinfo{journal}{~Phys. Rev. B} \textbf{\bibinfo{volume}{61}},
  \bibinfo{pages}{R16291} (\bibinfo{year}{2000}).

\bibitem[{\citenamefont{Sorella and Capriotti}(2010)}]{sorella_warp}
\bibinfo{author}{\bibfnamefont{S.}~\bibnamefont{Sorella}} \bibnamefont{and}
  \bibinfo{author}{\bibfnamefont{L.}~\bibnamefont{Capriotti}},
  \bibinfo{journal}{J. Chem. Phys.} \textbf{\bibinfo{volume}{133}},
  \bibinfo{pages}{234111} (\bibinfo{year}{2010}).

\end{thebibliography}

\end{document}



\widetext
\begin{center}
\textbf{\Large Supporting Material}
\vskip 0.3cm
\textbf{\large van der Waals forces stabilize low-energy polymorphism in B$_2$O$_3$:\\ implications for the {\em crystallization anomaly}} 
\vskip 0.2 cm
Guillaume Ferlat$^1$, Maria Hellgren$^1$, Fran\c cois-Xavier Coudert$^2$, Henri Hay$^1$, Francesco Mauri$^{1,3}$, Michele Casula$^1$
\vskip 0.1 cm
{\em $^1$Sorbonne Université, MNHN, UMR CNRS 7590, IRD, IMPMC, F-75005 Paris, France} \\
{\em $^2$Chimie ParisTech, PSL University, CNRS, Institut de Recherche de Chimie Paris, 75005 Paris, France} \\
{\em $^3$Dipartimento di Fisica, Universit\`a di Roma La Sapienza, Piazzale Aldo Moro 5, I-00185 Roma, Italy}
\end{center}
\setcounter{equation}{0}
\setcounter{figure}{0}
\setcounter{table}{0}
\setcounter{page}{1}
\makeatletter
\renewcommand{\theequation}{S\arabic{equation}}
\renewcommand{\thefigure}{S\arabic{figure}}
\renewcommand{\thetable}{S\arabic{table}}
\renewcommand{\bibnumfmt}[1]{[S#1]}
\renewcommand{\citenumfont}[1]{S#1}

\titlespacing*{\section}
{0pt}{3.0ex minus .2ex}{2.5ex plus .2ex}

\setlength{\floatsep}{5pt plus 2pt minus 2pt}
\setlength{\textfloatsep}{5pt plus 2pt minus 2pt}
\setlength{\intextsep}{5pt plus 2pt minus 2pt}

\section{B$_2$O$_3$ polymorphs}

A total of 27 B$_2$O$_3$ polymorphs (including B$_2$O$_3$-I) has been investigated: this includes the 26 structures originally considered in Ref.~\cite{ferl_2012_1} and a newly constructed one, derived from ideas suggested in Ref.~\cite{daub_2015_1} and whose structure is provided at the end of this document.
The set of structures of Ref.~\cite{ferl_2012_1} was obtained from a total of thirteen 3-connected nets (as derived from graph theory and originally applied to $sp^2$ carbon structures~\cite{wink_2001_1}) by decorating the nodes with either BO$_3$ triangles (Tx structures) or boroxol rings (Tx-$b$). For instance, T0 and T0-$b$ are the topological equivalent of graphite where carbon atoms have been substituted by BO$_3$ triangles and boroxols, respectively. 
Subsequently, an intercalated borosulfate (Cs$_3$H(B$_2$O$_3$)$_2$(SO$_4$)$_2$) was synthesized for the first time~\cite{daub_2015_1}. By removing the sulfate groups as well as the alkali and hydrogen atoms from this structure, one obtains a layered B$_2$O$_3$ compound whose motif (1 triangle and 1 boroxol) is intermediate between T0 and T0-$b$. We thus name this putative structure T0-0.5$b$ (we thank H. Hillebrecht for sending us an input structure). 
  
\section{Density functional theory (DFT)}

Total-energy calculations using the PBE~\cite{perd_96_1}, PBE+D2~\cite{grim_2006_1} and PBE+TS~\cite{tkat_2009_1} functionals were carried out with the CASTEP code~\cite{clar_2005_1}, while LDA, DF2~\cite{lee_2010_1} and DF-cx~\cite{thon_2015_1} computations used the Quantum ESPRESSO package~\cite{QE-2009}. The DF2 and DF-cx results are very close to each other and therefore the formers have been omitted from this report (except in table~\ref{table_energies} and figure~S2). 
For some of the networks, several energy minima - corresponding to different symmetries - were found. Only the lowest minimum of each topologically unrelated structure is reported.

As compared to Ref.~\cite{ferl_2012_1}, the CASTEP calculations were repeated with tighter (more accurate) ultra-soft pseudo-potentials, generated on-the-fly, and a larger plane-wave basis-set cutoff of 784 eV. This resulted only in marginal differences. 
However, for some topologies (e.g. T10-$b$), lower-energy and higher-density structures were found by using as input for the PBE calculations the output structures obtained from other functionals (note also that in Ref.~\cite{ferl_2012_1} an overall rescaling factor of 1.10 was applied on all the polymorphs densities to account for the estimated PBE error, as derived from the case of B$_2$O$_3$-I). 
We used a $k$-point grid spacing of 2$\pi\times$ 0.05 \AA$^{-1}$ or finer. Tolerances on the atomic forces and stress were set to 0.005 eV/\AA \ and 0.01 GPa, respectively.
Calculations with Quantum ESPRESSO also use ultra-soft pseudo-potentials~\cite{vand_90_1} albeit constructed differently and used with a higher energy cut-off of 1020 eV. Despite those differences and others inherent to the use of different codes, we checked explicitly (using the PBE functional) that both schemes provide essentially identical results, as can be judged from Fig.~S1.
\vskip -0.5cm
\begin{figure}[h!]
\centering
\includegraphics[width=0.52\columnwidth]{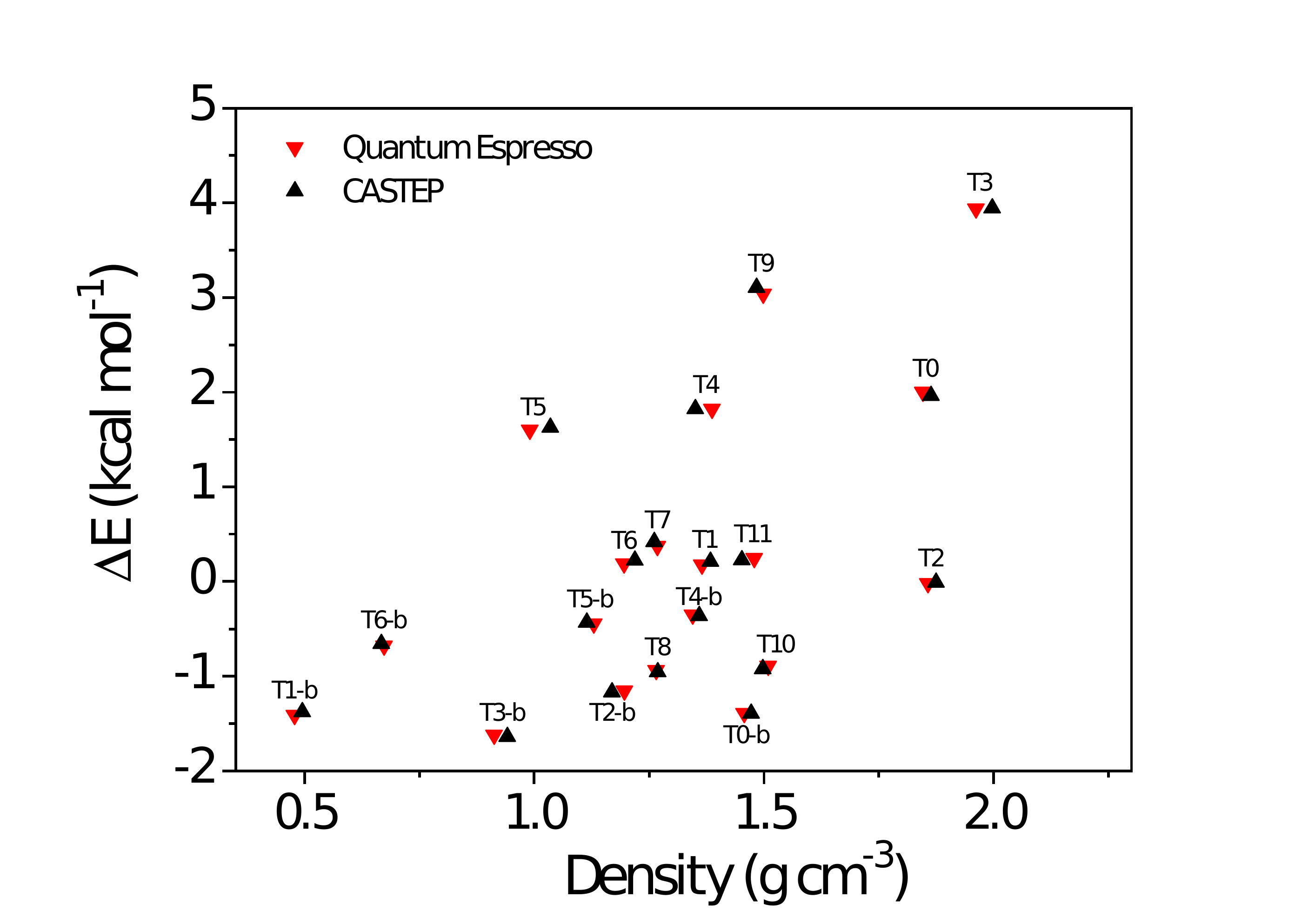}
\vskip -0.4cm
\caption{Comparison of the PBE results obtained with Quantum Espresso and CASTEP for several of the polymorphs.}
\end{figure}

Phonon frequencies ($\omega_i$) were computed using density functional perturbation theory, as implemented in the PHONON executable of the Quantum ESPRESSO package. 

\begin{figure*}
\includegraphics[width=1.0\textwidth]{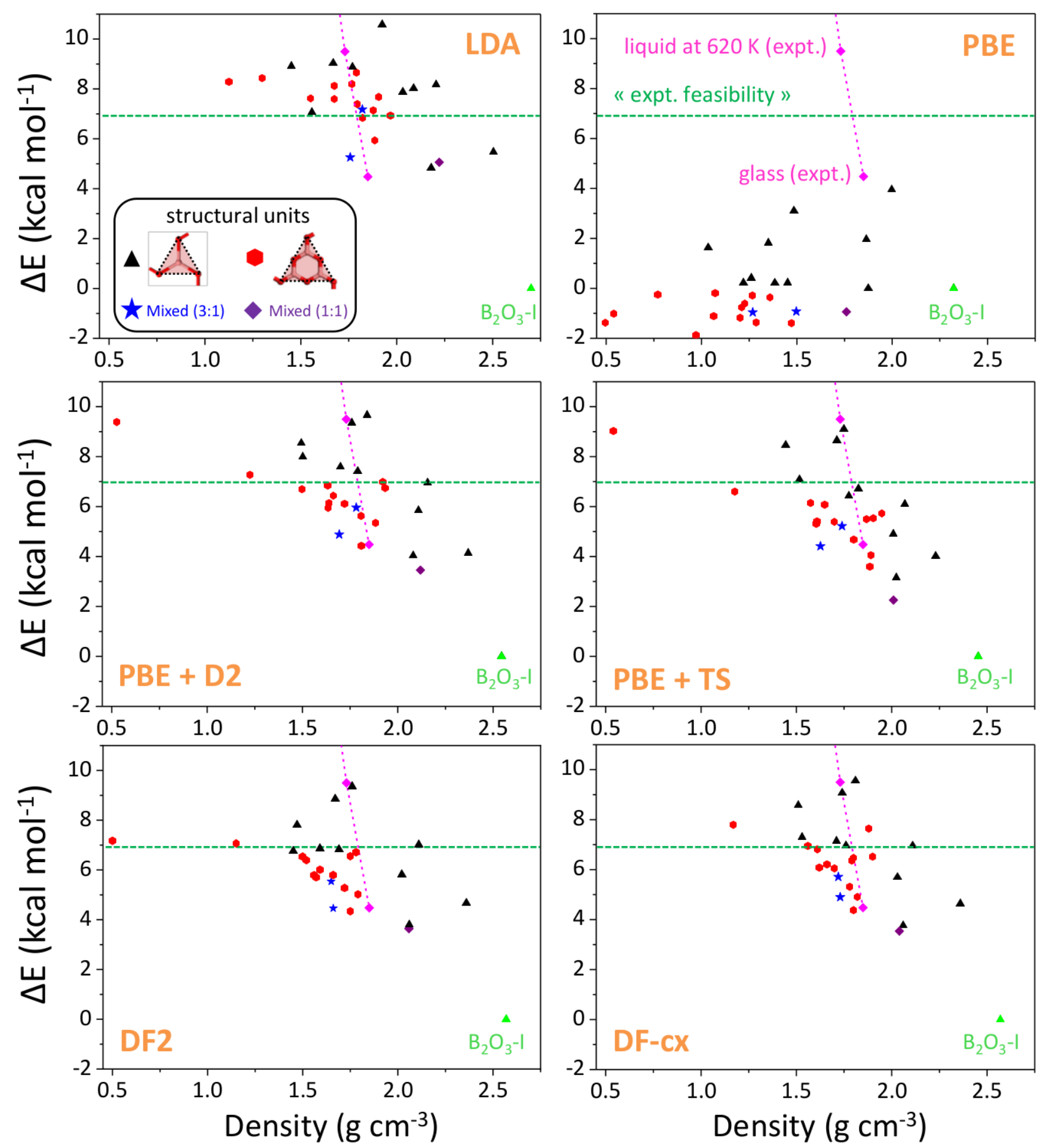}
\caption{
Energy-density for all polymorphs as obtained from the different DFT schemes. All energies are relative to that of ${\rm B_2O_3}$-I. 
The different symbols refer to the relative proportion of structural units (triangles:boroxol) in the polymorphs: $\blacktriangle$ only BO$_3$ triangles (1:0), \protect\hexa \ only boroxol rings (0:1), ${\color{blue} \bigstar}$ mixed decoration (3:1), \protect\lozen \ mixed decoration (1:1). 
The experimental energies for the liquid and glass states (\protect\loz) are from Ref.~\protect\cite{shmi_66_1}.
The horizontal green line is the estimated upper limit of metastability.
}
\label{fig_all_DFT}
\end{figure*}

From these frequencies, the zero-point energy (ZPE) contribution is directly given by $\displaystyle E_{vib} = \sum_i \frac{\hbar \omega_i}{2}$, and the vibrational free-energy at finite temperature, in the harmonic approximation, by $\displaystyle F_{vib}= k_bT \sum_i \ln \left[2 \sinh \left(\frac{\hbar \omega_i}{k_bT}\right)\right]$.
The calculations were carried out with LDA and PBE for all crystals and with PBE-D2 for a few crystals (Ref.~\cite{hay_thesis} and Tables \ref{table_ZPE}, \ref{table_Fvib}): the results show only weak dependencies upon the functional (less than 0.2 kcal~mol$^{-1}$).
Expressed relatively to that of B$_2$O$_3$-I, the ZPE differences ($\Delta E_{vib}$ at T = 0 K) are less than 0.6 kcal~mol$^{-1}$ (table \ref{table_ZPE}). At T = 300 K, the vibrational contributions ($\Delta F_{vib}$) are down-shifted for all crystals (including B$_2$O$_3$-I) by about $\sim$ 2-3 kcal~mol$^{-1}$, resulting in relative differences typically less than 1 kcal~mol$^{-1}$ (table \ref{table_Fvib}). Since the latter value is below the DFT accuracy, we did not attempt to include the vibrational contributions in the results shown in Figs.~7 and \ref{fig_all_DFT}.

\vskip -0.1cm
\begin{table}[h!]
\caption{Zero point energies (relative to B$_2$O$_3$-I, in kcal~mol$^{-1}$) obtained with the PBE functional.}
\label{table_ZPE}
\begin{tabular}{l c c c c c c c c c c c c c c}
\hline\hline
 polymorph          &B$_2$O$_3$-I& T0 & T1  & T2  & T3 & T4 & T5 & T6 & T7 & T8 & T9 & T10 & T11 & T0-0.5$b$\\
\hline
$\Delta E_{vib}$ \ (T = 0 \ K) & 0.0 &-0.3&-0.5&-0.2&0.0&0.0&0.0&0.0&0.0&0.0&-0.3&0.0&-0.3&0.0\\
 polymorph & B$_2$O$_3$-I-$b$ & T0-$b$ & T1-$b$ & T2-$b$ & T3-$b$ & T4-$b$ & T5-$b$ & T6-$b$ & T7-$b$ & T8-$b$ & T9-$b$ & T10-$b$ & T11-$b$ & \\
\hline
$\Delta E_{vib}$ \ (T = 0 \ K) &-0.3&-0.2&-0.3&-0.5&-0.3&0.2&-0.1&-0.1&-0.3&0.1&-0.2&-0.6&0.1& \\
\hline\hline
\end{tabular}
\end{table}
\vskip 0.2cm

\begin{table}[h!]
\caption{ZPE (T = 0 K) and vibrational free-energies (T = 300 K), in kcal~mol$^{-1}$, obtained with the PBE-D2 functional. The reference energy is the ZPE of B$_2$O$_3$-I.}
\label{table_Fvib}
\begin{tabular}{l c c c c c c c c c}
\hline\hline
 polymorph  &B$_2$O$_3$-I& T0 & T1 & T2 & T3 & T4 & T5 & T1-$b$ & T0-0.5$b$\\
\hline
$\Delta E_{vib}$ \ (T = 0 \ K) &  0.0 & -0.1 & 0.0 & 0.0 & -0.1 & -0.1 & -0.1 & -0.1 & -0.1 \\
$\Delta F_{vib}$ \ (T = 300 \ K) & -1.8 & -2.2 & -2.5 &-2.2 & -2.2 & -2.4 & -2.4 & -2.7 & -2.3\\
\hline\hline
\end{tabular}
\end{table}

The energy-density diagrams obtained for the full set of B$_2$O$_3$ polymorphs using the various DFT schemes, are reported in Fig.~\ref{fig_all_DFT}. The upper limit of metastability (horizontal green line) is estimated at 7 kcal/(mol B$_2$O$_3$) in analogy to SiO$_2$, a close parent system. 
Indeed, i) in both systems the glass energies - expressed in kcal/(mol of network former) - are equivalent within error bars: g-B$_2$O$_3$ is 2.2 kcal/(mol Boron) above B$_2$O$_3$-I~\cite{shmi_66_1} while g-SiO$_2$ is 1.9 $\pm$ 0.3 kcal/(mol Silicon) above quartz~\cite{rich_82_1}) ii) among low-density SiO$_2$ polymorphs (for which calorimetric data are available), the highest energy above quartz is 3.44 kcal/(mol SiO$_2$) (ISV zeolite)~\cite{picc_2000_1} which would correspond to 6.88 kcal/(mol B$_2$O$_3$) (the doubling comes from the number of network formers involved in the formula units). 

\section{Elastic properties}

Elastic properties were calculated at the DFT level, using the CRYSTAL14 code~\cite{dove_2014_1} allowing for a full account of the symmetry operators of each crystal according to its space group, thus reducing the computational power required. For each structure studied, we calculated its full tensor of second-order elastic constants through DFT calculations on a series of structures with small strains applied (±1\%). From these stiffness matrices, containing from 3 to 21 independent elastic constants depending on the Laue class of the crystal~\cite{mouh_2014_1} we performed tensorial analysis with the ELATE code~\cite{gail_2016_1} (available online at \url{http://progs.coudert.name/elate}) to extract the directional Young's modulus, shear modulus, linear compressibility and Poisson's ratio in the elastic regime. We also computed spatially averaged quantities following the Voigt--Reuss--Hill approximation. Representative input files for these calculations are available online at \url{https://github.com/fxcoudert/citable-data}.

\section{Random Phase Approximation (RPA)}
RPA total energies were calculated on top of PBE orbitals using the VASP code~\cite{kres_96_1} within the projector-augmented-wave PAW method~\cite{kres_99_1}. Calculations were performed on the structures as obtained from the DFT-TS functional, but we verified that energy differences were marginally sensitive to the choice of input structure. For example, a difference of less than 0.05 kcal/mol was found on T0-0.5$b$ with the DFT-D2 structure.

The correlation energy was converged with a 4x4x4 $k$-point grid (2x2x2 for T0-$b$ and T10) and Fock exchange on a 8x8x8 grid. With an energy cut-off of 600 eV energy differences were converged within 0.5 kcal/mol. 
The RPA energies obtained for a set of 5 polymorphs are reported in the main text (Figs. 4-7) and in Table~\ref{table_energies}.

\section{Quantum Monte Carlo (QMC) calculations}

We carried out QMC calculations by using the TurboRVB package~\cite{sorella_rvb}. The calculations are performed for an \emph{ab initio} Hamiltonian where the core electrons of oxygen and boron atoms have been replaced by scalar-relativistic pseudo-potentials of Burkatzki-Filippi-Dolg type~\cite{filippi_pseudo}, to reduce the computational cost. The QMC variational wavefunction is of Jastrow-Slater form, where both the Jastrow and the Slater determinant are expanded in Gaussian type orbitals (GTOs). The wavefunction has been extensively described elsewhere~\cite{Casula2004,michele_agp2,devaux2015electronic}. It provides an accurate description of ground state correlation. In particular, it is able to account for the van der Waals interactions, thanks to the three- and four-body Jastrow factors. 
By means of the QMC approach, we studied three B$_2$O$_3$ polymorphs, namely B$_2$O$_3$-I, T0, and T10, and computed energy differences and equilibrium geometries, to benchmark other methods. Since these polymorphs are near-degenerated in energy, a particular care is taken to converge the calculations with respect to relevant criteria, such as level of theory, finite-size effects, geometry and basis set.

As far as the level of theory is concerned, we performed both variational (VMC) and lattice-regularized diffusion Monte Carlo (LRDMC)~\cite{casula_diffusion_2005,filippi_lrdmc} calculations. Once the VMC wavefunction is optimized by energy minimization (we employed the linear method as energy minimization scheme~\cite{Umrigar2007}), the resulting variational form is further projected towards the ground state by means of the LRDMC algorithm in the fixed-node approximation.

\begin{figure}[h]%
\centering
\includegraphics[width=0.6
\columnwidth]{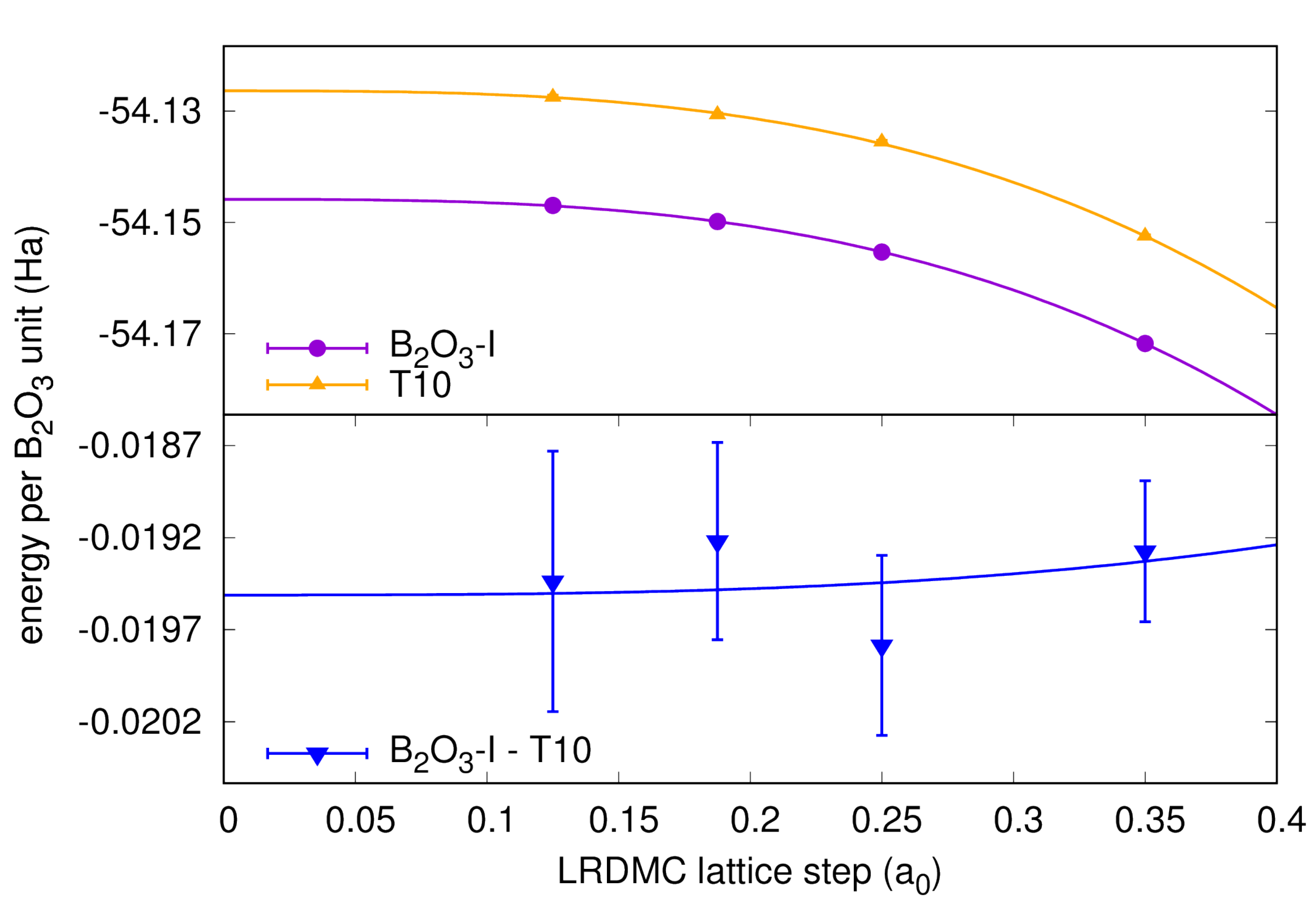}
\vskip -0.3cm
\caption{LRDMC lattice step extrapolation for the B$_2$O$_3$-I and T10 polymorphs, in the smallest orthorhombic supercell fitting the unit cell motif. Upper panel: the total energy per B$_2$O$_3$ unit is plotted as a function of the lattice step $a$ for both polymorphs. The fitting functions are of the form $c_0 + c_2 a^2 + c_3 a^3$. Lower panel: the energy difference per B$_2$O$_3$ unit between  B$_2$O$_3$-I and T10 is plotted for each $a$. In the energy differences, the convergence on $a$ is much faster. We used $a=0.125 ~a.u.$ for all other LRDMC calculations, which yields safely converged results. \label{LRDMC_lattice_space_extrapolation}}%
\end{figure}

The fixed-node projection provides nearly exact energy differences. 
In the LRDMC procedure, we worked with lattice step ($a=0.125$ a.u.), branching time ($\tau=0.01$ a.u.), and walker population size (population fixed at $N_w=1024$ walkers~\cite{Calandra1998}), that yield unbiased energy estimates. We project the wavefunction for a total imaginary time evolution of about 15 a.u., which generates a sample large enough to get statistical error bars of less than 0.5 kcal/(mol B$_2$O$_3$). The LRDMC lattice step convergence test is reported in Fig.~\ref{LRDMC_lattice_space_extrapolation}.

\vskip -0.1cm

One of the major sources of bias, present in any quantum Monte Carlo calculation of extended systems, comes from the finite-size dependence of the QMC energy estimates. 
The requested accuracy is $< 0.5$ kcal/(mol B$_2$O$_3$), and we adopt several strategies to achieve this target. The first one is to take into account orthorhombic supercells compatible with the lattice motif of a given polymorph. Orthorhombic supercells have usually a faster convergence with the system size, because the related electronic configuration is a closed shell. Second, for each orthorhombic supercell, we use the special $k$-point extrapolation~\cite{dagrada2016exact}, where the wavefunction is taken at the $k$-point whose DFT energy is equal to the one averaged over the whole Brillouin zone (the DFT converged energy). The $k$-point is determined at the LDA level, given the supercell and crystal symmetry. The special $k$-point boundary conditions cure the so-called one-body size effects. Third, we correct our energies for the size effects arising from the long-range part of the interacting Hamiltonian (the two-body size effects), by taking Hartree-Fock (HF) as the interacting reference theory.

\begin{figure}[h]%
\begin{center}
\includegraphics[width=0.45\columnwidth]{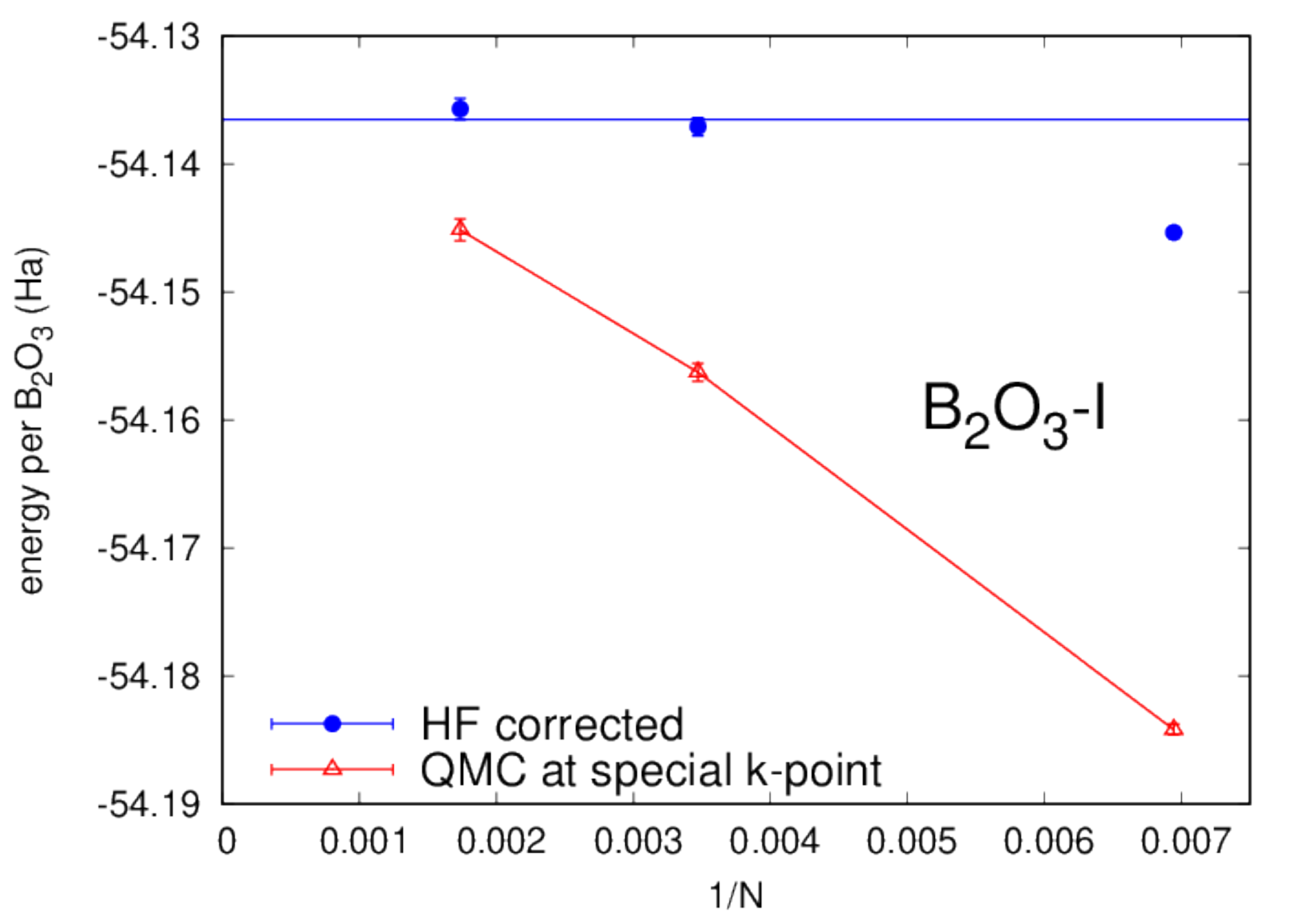}
\includegraphics[width=0.45\columnwidth]{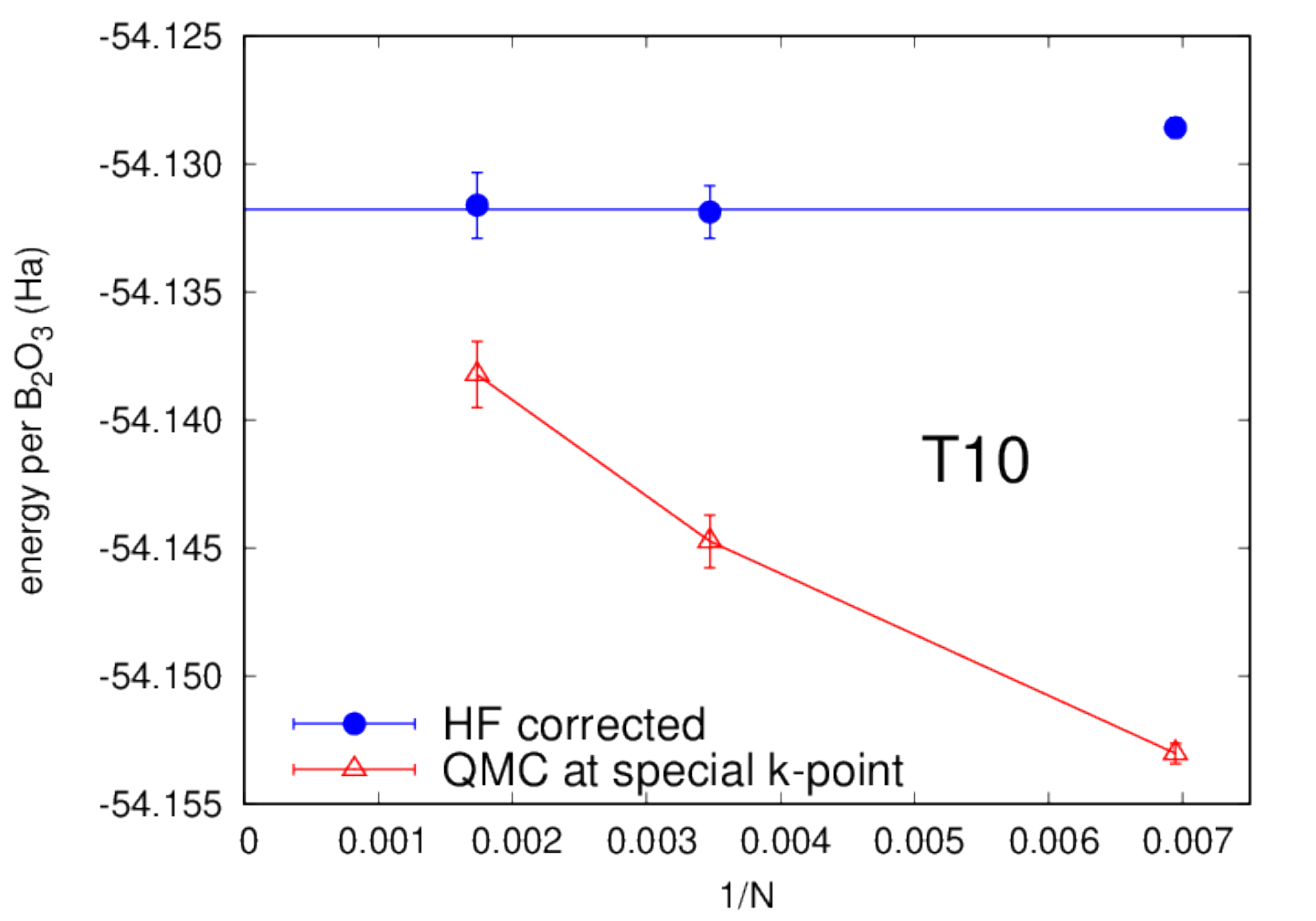}%
\end{center}
\vskip -0.3cm
\centering
\includegraphics[width=0.5
\columnwidth]{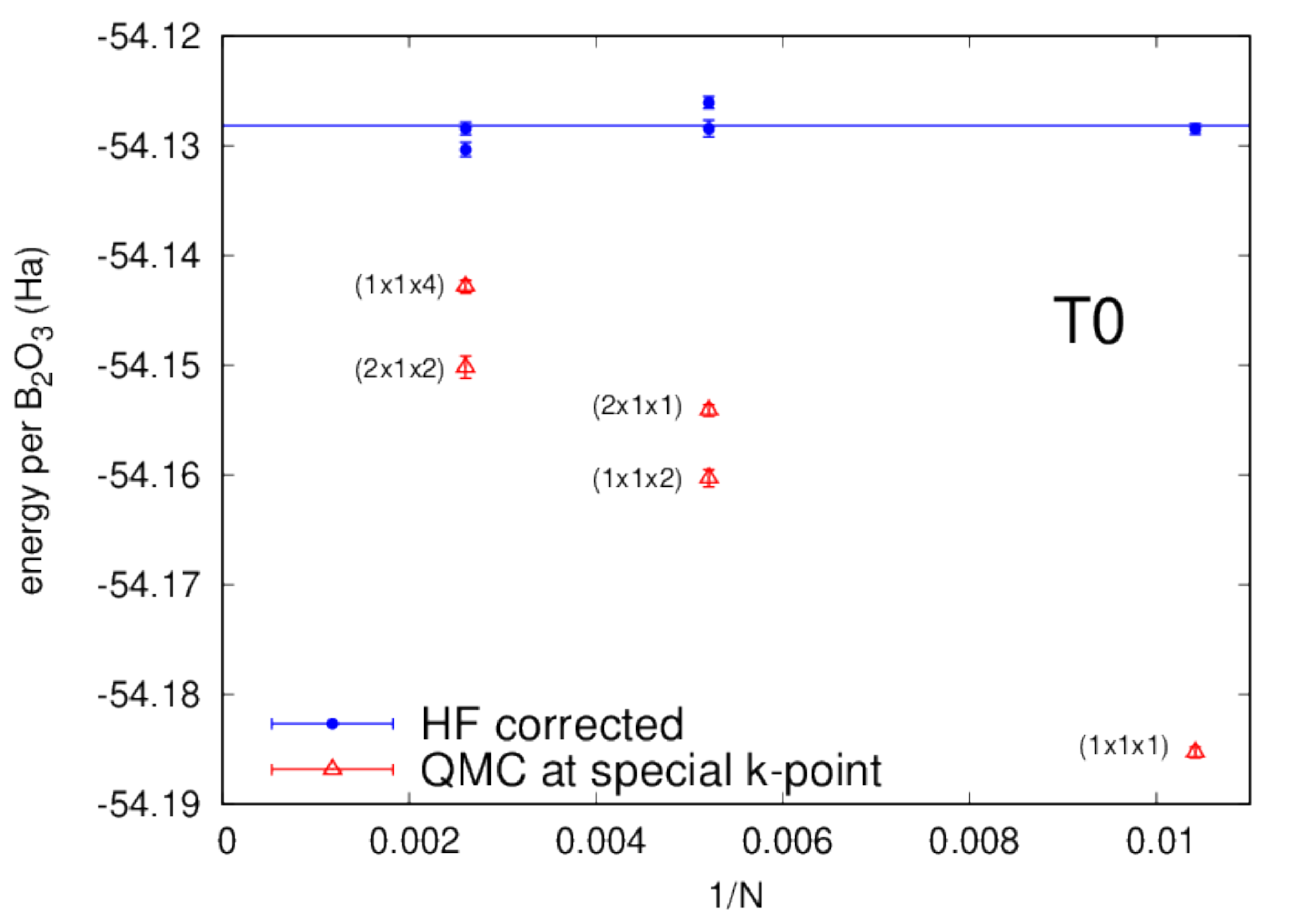}
\vskip -0.3cm
\caption{QMC finite-size extrapolation for the B$_2$O$_3$-I (top left panel), T10 (top right panel), and T0 (bottom panel) crystals. The bare special $k$-point calculations are shown, together with the extrapolations corrected by HF as reference theory. 
The most difficult extrapolation is for the T0 polymorph, due to its high anisotropy.
\label{QMC_extrapolations}}%
\end{figure}

The actual expression for the size-corrected energy is the following:
\begin{equation}
E^{QMC}_\textrm{corr}(L) = E^{QMC}_{k_s}(L) + 2\left(E^{HF}(\infty) - E^{HF}_{k_s}(L)\right),
\label{HFcorrection}
\end{equation}
where $k_s$ is the special $k$-point, $E^{QMC}_{k_s}(L)$ is the QMC energy at the given size $L$ computed at $k_s$, $E^{HF}_{k_s}(L)$ is the HF energy computed on a $k$-point grid corresponding to the supercell size $L$, shifted by $k_s$, while $E^{HF}(\infty)$ is the HF energy converged at the thermodynamic limit. The difference $E^{HF}(\infty) - E^{HF}_{k_s}(L)$ in the right-hand side of Eq.~\ref{HFcorrection} mainly takes into account long-range Coulomb interaction effects, as the finite-size calculations are performed at the special $k$-point, which exactly cancels out one-body effects by definition. We ran the VASP code to carry out the HF calculations~\cite{kres_96_1}. 
In order to perform the energy extrapolation as a function of $1/L$ (or equivalently as a function of $1/N$), we took into account several cell sizes for each polymorph. For B$_2$O$_3$-I and T10, we used $L=\{(1\times 1\times 1), (2\times 1 \times 1), (2\times 2 \times 1)\}$, where the smallest supercell $L=(1\times 1\times 1)$ is the orthorhombic one containing 144 electrons. For T0, we studied five supercells $L=\{(1\times 1\times 1), (2\times 1 \times 1), (1\times 1 \times 2), (2\times 1 \times 2), (1\times 1 \times 4)\}$, with the smallest orthorhombic one containing 96 electrons (and one layer, with staking in the $z$ direction). The results of the energy extrapolations are shown in Fig.~\ref{QMC_extrapolations}. The HF-corrected energies provide a significant improvement in the energy extrapolation, with respect to the sole special $k$-point scaling. 

It is of paramount importance to find out the equilibrium structural properties of the B$_2$O$_3$ polymorphs, especially in the case of the predicted structures (for which experimental data are not available). 
Within the QMC framework, geometry relaxation has become feasible recently, with the development of improved (finite variance) force estimators~\cite{Casula2004,estim1,umrigar_warp}, and efficient methods 
such as the algorithmic differentiation~\cite{sorella_warp,barborini2012structural}. For computational cost reasons, the geometry relaxation is doable in the VMC framework only. We carried out geometry optimization in supercells of moderate-large size, i.e. in the $(2\times 1 \times 1)$ cell for both B$_2$O$_3$-I and T10 and in the $(2\times 1 \times 2)$ cell for T0, by computing the electronic structure at the special $k$-point. In the VMC geometry relaxation scheme, both ionic configurations and electronic parameters are evolved to minimize the total (electrons+nuclei) variational energy. 

To make the geometry optimizations computationally feasible, we used a not-so-large basis set, with the aim at reducing the total number of variational wavefunction parameters. We derived a minimal basis set for both O and B by optimizing the coefficients and the exponents of the double $\zeta$ basis set of Burzatki-Filippi-Dolg~\cite{filippi_pseudo} at the VMC level, and getting rid of the orbitals with the lowest coefficients. We then used the resulting primitive GTOs to construct the contracted basis set (6s4p)/[2s1p] for B, and (5s4p)/[2s1p] for O. The molecular orbitals that enter the Slater determinant are initialized using the LDA orbitals developed on the same basis set, before being optimized at the QMC level. The Jastrow part of the wavefunction contains a homogeneous two-body factor, made of a simple Pad\'e parametrization, and three- and four-body terms, expanded on a $2s2p1d$ GTO basis set. After the geometry relaxation, we performed final VMC and LRDMC calculations at frozen equilibrium geometry with a large uncontracted basis set for the determinant, made of (10s10p1d) GTOs for both B and O, with Gaussian exponents previously optimized at the LDA level. In this last step, the molecular orbitals generated by LDA are kept frozen in QMC. We verified that the LDA orbitals, developed on our large primitive basis set, generate optimal nodes. Indeed, further orbital optimization at the VMC level does not improve the LRDMC energy. The three- and four-body Jastrow factors are developed on a $2s2p1d$ GTO basis set. These calculations, performed at the optimal geometries and at the converged basis set, are then extrapolated to the thermodynamic limit, with independent extrapolations for each polymorph reported in Fig.~\ref{QMC_extrapolations}, to yield the final energy differences (Tab.~\ref{table_energies}). 


\section{Energy differencies for a set of B$_2$O$_3$ polymorphs computed across different methods}



\begin{table*}[h!]
\caption{
Energies, relative to B$_2$O$_3$-I (in kcal~mol$^{-1}$) obtained from the different {\em ab initio} schemes for a set of 5 polymorphs.}
\label{table_energies}
\begin{tabular}{l c c c c c c c c}
\hline\hline
          & LDA &  PBE & PBE+D2 & PBE+TS & DF2 & DF-cx & RPA & QMC    \\ 
\hline
T0        & 5.5 &  2.0 &  4.1   &  4.0   & 4.7 &  4.6  & 4.2 & 4.7 $\pm$ 0.9 \\ 
T0-0.5$b$ & 5.1 & -0.9 &  3.5   &  2.2   & 3.7 &  3.5  & 1.9 & - \\
T0-$b$    & 5.9 & -1.4 &  4.4   &  3.6   & 4.3 &  4.4  & 2.5 & - \\
T3        & 8.2 &  3.9 &  7.0   &  6.1   & 7.0 &  6.9  & 7.2 & - \\
T10       & 5.3 & -0.9 &  4.9   &  4.4   & 4.5 &  4.9  & 3.4 & 3.0 $\pm$ 0.4 \\ 
\hline\hline
\end{tabular}
\end{table*}

\clearpage

\section{Structure of selected B$_2$O$_3$ polymorphs}

See Table S4 for the structure of the novel T0-0.5$b$ polymorph, as relaxed by the PBE+D2 scheme, and Table S5 for the structure of 2 polymorphs (T0 and T10), as relaxed by the VMC scheme.

\setlength{\tabcolsep}{9pt}
\begin{table*}[h!]
\caption[]{Number of ${\rm B_2O_3}$ units per unit cell (N), space group (SG), lattice parameters (\AA), cell angles and atomic positions.}
\label{table_struc}
\begin{tabular}{c c c c c c c c c c}
%
 Polymorph &  N &   SG       & \multicolumn{3}{c}{Lattice parameters} & \multicolumn{4}{c}{Atom positions}    \\
           &    &            & \multicolumn{3}{c}{and cell angles}    & Atom &  $x/a$   &   $y/b$  &   $z/c$  \\
\hline
\hline
T0-0.5$b$  &  4 & $P$63/m    & \ 6.627 & \ 6.627 & \  5.740           &  B   &  0.0000  &  0.0000  &  0.2500  \\
           &    &            &   90.00 &  90.00  &  120.00            &  B   &  0.2655  &  0.4335  &  0.2500  \\
           &    &            &         &         &                    &  O   &  0.0990  &  0.5001  &  0.2500  \\
           &    &            &         &         &                    &  O   &  0.2095  &  0.2045  &  0.2500  \\
\hline
\end{tabular}
\end{table*}


\begin{table*}[h!]
\caption[]{Number of ${\rm B_2O_3}$ units per unit cell (N), space group (SG), lattice parameters (\AA), cell angles and atomic positions.}
\label{table_struc_QMC}
\begin{tabular}{c c c c c c c c c c}
%
 Polymorph &  N &   SG       & \multicolumn{3}{c}{Lattice parameters} & \multicolumn{4}{c}{Atom positions}    \\
           &    &            & \multicolumn{3}{c}{and cell angles}    & Atom &  $x/a$   &   $y/b$  &   $z/c$  \\
\hline
\hline
    T0     &  2 & $P$2$_1$/m &  4.348  &  5.631  &    4.348        &  B   &  0.1779  &  0.2500  &  0.0256  \\
           &    &            &  90.00  &  120.00 &   90.00         &  B   &  0.8407  &  0.2500  &  0.3575  \\
           &    &            &         &         &                 &  O   &  0.1295  &  0.2500  &  0.6905  \\
           &    &            &         &         &                 &  O   &  0.8900  &  0.2500  &  0.0704  \\
           &    &            &         &         &                 &  O   &  0.5000  &  0.2500  &  0.3102  \\
\hline
   T10     & 6  &  $P$1      &  4.330  &  6.556  & 14.910          &  B   &  0.7136  &  0.0300  &  0.8642  \\
           &    &            &  90.00 &   90.00  &  90.00          &  B   &  0.3769  &  0.9391  &  0.9967  \\
           &    &            &         &         &                 &  B   &  0.8775  &  0.9385  &  0.0821  \\
           &    &            &         &         &                 &  B   &  0.2349  &  0.9838  &  0.2162  \\
           &    &            &         &         &                 &  B   &  0.5426  &  0.2283  &  0.2917  \\
           &    &            &         &         &                 &  B   &  0.7875  &  0.5310  &  0.3644  \\
           &    &            &         &         &                 &  B   &  0.1249  &  0.4399  &  0.4965  \\
           &    &            &         &         &                 &  B   &  0.6246  &  0.4395  &  0.5820  \\
           &    &            &         &         &                 &  B   &  0.2654  &  0.4839  &  0.7158  \\
           &    &            &         &         &                 &  B   &  0.9558  &  0.7280  &  0.7913  \\
           &    &            &         &         &                 &  B   &  0.5912  &  0.8761  &  0.3256  \\
           &    &            &         &         &                 &  B   &  0.9070  &  0.3761  &  0.8251  \\
           &    &            &         &         &                 &  O   &  0.4892  &  0.9390  &  0.9131  \\
           &    &            &         &         &                 &  O   &  0.8780  &  0.9242  &  0.8032  \\
           &    &            &         &         &                 &  O   &  0.7486  &  0.2345  &  0.8721  \\
           &    &            &         &         &                 &  O   &  0.0664  &  0.9362  &  0.0095  \\
           &    &            &         &         &                 &  O   &  0.9824  &  0.9393  &  0.1666  \\
           &    &            &         &         &                 &  O   &  0.3289  &  0.1801  &  0.2276  \\
           &    &            &         &         &                 &  O   &  0.3644  &  0.8344  &  0.2665  \\
           &    &            &         &         &                 &  O   &  0.6221  &  0.4246  &  0.3037  \\
           &    &            &         &         &                 &  O   &  0.6639  &  0.0757  &  0.3419  \\
           &    &            &         &         &                 &  O   &  0.0114  &  0.4396  &  0.4131  \\
           &    &            &         &         &                 &  O   &  0.7501  &  0.7352  &  0.3725  \\
           &    &            &         &         &                 &  O   &  0.4355  &  0.4368  &  0.5093  \\
           &    &            &         &         &                 &  O   &  0.5169  &  0.4393  &  0.6663  \\
           &    &            &         &         &                 &  O   &  0.1702  &  0.6797  &  0.7269  \\
           &    &            &         &         &                 &  O   &  0.1337  &  0.3345  &  0.7658  \\
           &    &            &         &         &                 &  O   &  0.8346  &  0.5754  &  0.8413  \\
           &    &            &         &         &                 &  O   &  0.5650  &  0.9369  &  0.0698  \\
           &    &            &         &         &                 &  O   &  0.9362  &  0.4382  &  0.5695  \\
\hline
\end{tabular}
\end{table*}




